\documentclass[aps]{revtex4-2}
\usepackage{graphicx}
\usepackage{color}
\def\no{\noindent}
\def\bc{\begin{center}}
\def\ec{\end{center}}

\def\beq{\begin{equation}}
\def\eeq{\end{equation}}

\def\br{{\bf r}}

\begin{document}

\title{
Randomly repeated measurements on quantum systems:\\
Correlations and topological invariants of the quantum evolution
}

\author{K. Ziegler$^1$, E. Barkai$^2$ and D. Kessler$^3$
}
\affiliation{$^1$Institut f\"ur Physik, Universit\"at Augsburg, D-86135 Augsburg, Germany\\
$^2$Department of Physics, Institute of Nanotechnology and Advanced Materials, 
Bar-Ilan University, Ramat Gan 52900, Israel\\
$^3$Department of Physics, Bar-Ilan University, Ramat Gan 52900, Israel
}
\date{\today}

\begin{abstract}
Randomly repeated measurements during the evolution of a closed quantum system
create a sequence of probabilities for the first detection of a certain quantum state.
The related discrete monitored evolution for the return of the quantum system
to its initial state is investigated. We found that the mean number of measurements
until the first detection is an integer, namely the dimensionality of 
the accessible Hilbert space. Moreover, the mean first detected return time is equal to
the average time step between successive measurements times the mean number of measurements.
Thus, the mean first detected return time scales linearly with the dimensionality of the
accessible Hilbert space. The main goal of this work is to explain the quantization
of the mean return time in terms of a quantized Berry phase.
\end{abstract}

\maketitle

\section{Introduction}

The unitary evolution (UE) of a closed quantum system from the initial state $|\Psi\rangle$ to the state 
$|\Psi(\tau)\rangle$ on the time interval $\tau$ is defined by $|\Psi(\tau)\rangle=\exp(-iH\tau/\hbar)|\Psi\rangle$,
where $H$ is the Hamiltonian of the system. The result of this
evolution is characterized by the overlap amplitude $\langle\psi|\Psi(\tau)\rangle$ with respect to a given state $|\psi\rangle$.
Then $|\langle\psi|\Psi(\tau)\rangle|^2$ is the probability that the evolution has reached the state $|\psi\rangle$,
which obviously depends on the time $\tau$. 
In a single experiment the measurement time is fixed and, therefore, the UE
allows us to detect the state $|\psi\rangle$ with probability $|\langle\psi|\Psi(\tau)\rangle|^2$ only once. 
The detection of this state at different times
would require the repetition of the experiment, prepared in the same initial state $|\Psi\rangle$, for different values of 
$\tau$~\cite{Stefanak2008}.
An alternative approach, which we call ``monitored evolution (ME)'' is to allow the system to evolve from the initial state 
$|\Psi\rangle$ for the time
$\tau_1$ and then measure whether or not the system is in the state $|\psi\rangle$. If the answer is ``yes'', we stop
the experiment, if the answer is ``no'' we allow the system to evolve further after the measurement~\cite{bach04}
and perform a second measurement at time $\tau_1+\tau_2$. This procedure is repeated for times 
$t_2=\tau_1+\tau_2$, ..., $t_k=\tau_1+\cdots +\tau_k$ until the measurement detects the state $|\psi\rangle$ for the first time. 
We consider cases in which the state is detected with probability one. This is called a recurrent measurement process 
\cite{Stefanak2008}.
If at time $t_1$ the outcome of the measurement is null this
can be associated with a projection of the quantum system $({\bf 1}-|\psi\rangle\langle\psi|)|\Psi(t_1)\rangle$ with a 
subsequent normalization of the resulting state~\cite{Friedman2017}, and similarly for $t_2$ etc.
And when the state is detected for the first time at time $t_k$, the amplitude of the corresponding state is
$\phi_k$ and its probability is $|\phi_k|^2$ \cite{dhar15,Dhar_2015}.

For fixed time steps $\tau$ with $t_k=k\tau$ the repeated measurement approach, also known as the stroboscopic protocol, 
has been studied in great detail in Refs. \cite{Stefanak2008} and 
\cite{krovi06,krovi06a,krovi07,Gruenbaum2013,luck14,Grunbaum2014,dhar15,Dhar_2015,sinkovicz16,Friedman_2016,
Friedman2017,thiel18,nitsche18,2018JSMTE..02.3104K,lahiri19,Yin2019,2020arXiv201201196D}. 
Two cases have been distinguished, the return
probability for $|\psi\rangle=|\Psi\rangle$ and the transition probability for $|\psi\rangle\ne|\Psi\rangle$.
It was found that 
(i) the return and the transition probabilities differ qualitatively,
(ii) the average first detected return time is quantized and given by the winding number of the Laplace transform of
the return amplitude~\cite{Stefanak2008}, 
(iii) near degeneracies of the spectrum of the evolution operator, the fluctuations become very large and diverge at the 
degeneracies~\cite{Grunbaum2014,Friedman2017,Yin2019},
and (iv) the average first detected transition time also diverges at the degeneracies~\cite{quancheng20}.

Our main intention is to present in this article a study of the effect of independent and identically
distributed random time steps $\{\tau_k\}$ on the ME. For this purpose we want to answer the following questions:
(1) Do we still obtain quantization of the number of attempts for the first successful measurement?
(2) Is the average mean time for the first detection also quantized, as found for stroboscopic measurements?
(3) How can dynamical quantization be related to topological invariants?
(4) Do random time steps affect the divergent fluctuations near resonances?
To answer these questions we will develop a theory of the first detection time under repeated random measurements.
%
Our work is based on the ideas presented by Gr\"unbaum et al.~\cite{Gruenbaum2013} for fixed time steps and related
to a work by Varbanov et al.~\cite{varbanov08} on random time steps, who studied the conditions for the existence 
of non-detectable dark states.
Random time measurements were discussed also for open quantum systems \cite{2020arXiv201104403R}, while
for closed quantum systems they were recently studied by us in terms of average return and transition 
probabilities \cite{PhysRevA.103.022222}. The present work is an extension of the latter in which we explain in
detail the origin of the quantization of the mean return time. For this purpose we compare the Berry phase
of the return amplitude, averaged with respect to the distribution of the measurement times. It will be
shown that the average Berry phase is equal to the average number of measurements for the first detected return
and equal to the dimensionality of the accessible Hilbert space. The average Berry phase is reminiscent of the quantized
winding number in case of stroboscopic measurements \cite{Gruenbaum2013}.
   
The structure of this article is as follows. 
In Sect. \ref{sect:model} the definitions of the relevant quantities for the first detection under repeated 
measurements are given.
Then a short summary of the main results are presented in Sect. \ref{sect:summary}.
After a brief discussion of the ME for fixed time steps $\tau$ 
in Sect. \ref{sect:fixed_time}, we formulate the ME for random time steps in terms of random matrix products 
in Sects. \ref{sect:matrix}-\ref{sect:mean}. This includes a description of the averaging procedure 
(Sect. \ref{sect:time_av}) and the introduction of a generating function for the average return
time and its higher moments (Sect. \ref{sect:gen_funct}). The quantization of the mean number of measurements of the
first detected return by the dimensionality of the accessible Hilbert space is studied in Sects. \ref{sect:quantization1} 
and \ref{sect:mean}.
Finally, in Sect. \ref{sect:example} the example of a symmetric two-level system is analyzed, followed by a discussion of 
all the results in Sect. \ref{sect:discussion}.
Details of the calculations are presented in Apps. \ref{app:norm}-\ref{app:2ls}.

\section{Return amplitude of the monitored evolution}
\label{sect:model}

The return probability as a function of time, with or without intermediate measurements, provides a measure of how big the 
accessible space is and how long it takes to return to the initial state. This is an important 
quantity for classical random walks \cite{Redner2001,Bnichou2015} and plays also an important role 
for characterizing localization in many-body quantum systems~\cite{1987PhRvA..35.1360H,cohen16}. 
We will investigate the return amplitude for the ME with measurements at time steps $\{\tau_k\}_{k=1,2,...}$. 
First, we will have need to refer to the return amplitude of the UE
\beq
u_k=\langle\Psi|e^{-iH(\tau_1+\cdots+\tau_k)}|\Psi\rangle
\ ,
\label{ue_amp}
\eeq
for the state $|\Psi\rangle$ when we measure only once after the time $t_k=\tau_1+\cdots+\tau_k$,
assuming that the UE is governed by the Hamiltonian $H$. 
Then we turn to the return amplitude for the ME \cite{dhar15,Friedman_2016,Friedman2017}
\beq
\phi_{k}=\langle\Psi|e^{-iH\tau_k}(Pe^{-iH\tau_{k-1}})\cdots(Pe^{-iH\tau_1})|\Psi\rangle
\ ,\ \ \
P={\bf 1}-|\Psi\rangle\langle\Psi|
\ ,
\label{me_amp}
\eeq
which is the major objective of our inquiry.
This is the return amplitude at time $t_k$, provided that we also measure at times $t_1,t_2, ...,t_{k-1}$
but detect the quantum state $|\Psi\rangle$ for the first time at $t_k$ with probability $|\phi_k|^2$.  
For the time of measurements we assume a distribution density ${\cal P}(\{\tau_k\})$ of independent and identically 
distributed time steps, such that ${\cal P}(\{\tau_k\})=\prod_k P(\tau_k)$. 
This can be understood as the effect of an inaccurate clock. 
This enables us to consider the average $\langle ...\rangle_\tau\equiv\int ... \prod_k {\cal P}(\tau_k)d\tau_k $ 
with respect to the ensemble of random time steps. Here we do not specify the distribution.
For instance, ${\cal P}(\tau_k)$ can be a Dirac delta, which would recover the
stroboscopic ME \cite{Gruenbaum2013,Friedman_2016,Friedman2017,Yin2019}, it could be an exponential distribution 
or any other distribution, which allows us to perform the average $\langle ...\rangle_\tau$.

Besides the averaging with respect to times steps $\{\tau_k\}$ we also need to define the averages 
\beq
\overline{k^m}:=\sum_{k\ge 1}k^m\langle|\phi_k|^2\rangle_\tau
\ ,\ \ \ 
\overline{t^m}:=\sum_{k\ge 1}\langle t_k^m|\phi_k|^2\rangle_\tau
\ \ \ \ (m=1,2,...)
\ .
\label{overline}
\eeq
Here it is assumed that $\sum_k|\phi_k|^2=1$, which is justified for a finite-dimensional Hilbert space
(cf. App. \ref{app:norm}). This equation means that the state is eventually detected \cite{Gruenbaum2013}.  
Thus, the overline represents a double average, namely
an average with respect to the number of time steps $k=1,2,...$ with weight $|\phi_k|^2$, followed by 
the average $\langle ...\rangle_\tau$. 
For $m=1$ both expressions in Eq. (\ref{overline}) will be used to characterize the ME: $\overline{k}$
is the mean number of measurements for the first detected return (FDR) and $\overline{t}$ is the mean FDR time.


Since both evolutions are defined on an $N$-dimensional Hilbert space
by the Hamiltonian $H$, we consider its eigenstates $\{|E_j\rangle\}_{j=1,...,N}$ and its
corresponding eigenvalues $\{E_j\}_{j=1,...,N}$.
Then the return amplitude $\phi_k$ of the ME can be expressed as a matrix product, which is efficiently
written in the energy representation as a sum over all energy levels $\{E_j\}_{j=1,2,...,N}$ as
\beq
\phi_{k}=\sum_{j_1,j_2,...,j_k=1}^N
\langle\Psi|E_{j_k}\rangle e^{-iE_{j_k}\tau_k}\langle E_{j_k}|P|E_{j_{k-1}}\rangle e^{-iE_{j_{k-1}}\tau_{k-1}}
\cdots \langle E_{j_2}|P|E_{j_1}\rangle e^{-iE_{j_1}\tau_1}\langle E_{j_1}|\Psi\rangle
\ .
\label{matrix_proda}
\eeq
In principle, there is the possibility of degenerate eigenvalues or of a vanishing overlap 
$p_j=|\langle E_j|\Psi\rangle|^2$, which must be treated with care \cite{Gruenbaum2013}.
%
%
To understand the effect of degenerate energy levels on $\phi_k$ we consider the levels $E_j$, $E_{j'}$ and 
assume at first that they are not degenerate. Then we can write
\[
\sum_{j_k=1}^N\langle E_{j_{k-1}}|P|E_{j_{k}}\rangle e^{-iE_{j_{k}}\tau_{k}}\langle E_{j_{k}}|P|E_{j_{k+1}}\rangle
=\sum_{j_k=1,j_k\ne j,j'}^N e^{-iE_{j_{k}}\tau_{k}}\langle E_{j_{k-1}}|P|E_{j_{k}}\rangle\langle E_{j_{k}}|P|E_{j_{k+1}}\rangle
\]
\[
+\langle E_{j_{k-1}}|P[e^{-iE_{j}\tau_{k}}|E_{j}\rangle \langle E_{j}|+e^{-iE_{j'}\tau_{k}}
|E_{j'}\rangle\langle E_{j'}|]P|E_{j_{k+1}}\rangle
\ ,
\]
where the second term on the right-hand side describes the evolution of the state $P|E_{j_{k+1}}\rangle$
with the evolution operator
\[
P[e^{-iE_{j}\tau_{k}}|E_{j}\rangle \langle E_{j}|+e^{-iE_{j'}\tau_{k}}
|E_{j'}\rangle\langle E_{j'}|]
\]
over the time period $\tau_k$. The evolution creates a superposition of the states $|E_{j}\rangle$ and $|E_{j'}\rangle$,
which changes in time due to the time dependent coefficients, provided that $E_j\ne E_{j'}$. On the other
hand, for the degenerate case $E_j=E_{j'}$, the superposition of $|E_{j}\rangle$ and $|E_{j'}\rangle$
is fixed during the time period $\tau_k$:
\[
e^{-iE_{j}\tau_{k}}P[|E_{j}\rangle \langle E_{j}|+|E_{j'}\rangle\langle E_{j'}|]
\]
and only the global phase changes. This reflects a dimensional reduction of the accessible Hilbert space by 1,
implying that we should use the replacement
\[
|E_j\rangle, |E_{j'}\rangle \to |E_{jj'}\rangle
:=|E_{j}\rangle\langle E_j|\Psi\rangle +|E_{j'}\rangle\langle E_{j'}|\Psi\rangle
\]
and the simultaneous elimination of $j'$ from the summation of $j_k$ in Eq. (\ref{matrix_proda}).

Another special case is a vanishing overlap $p_j=|\langle E_j|\Psi\rangle|^2=0$.
Beginning with the initial state on the right-hand side of Eq. (\ref{matrix_proda}) we get
\[
\sum_{j_1=1}^N\langle E_{j_2}|P|E_{j_1}\rangle e^{-iE_{j_1}\tau_1}\langle E_{j_1}|\Psi\rangle
=\sum_{j_1=1;j_1\ne j}^N\langle E_{j_2}|P|E_{j_1}\rangle e^{-iE_{j_1}\tau_1}\langle E_{j_1}|\Psi\rangle
\ .
\]
Next, in the summation with respect to $j_2$ the special value $j$ does not contribute again, 
since $j_2=j$ gives
\[
\langle E_{j}|P|E_{j_1}\rangle 
= \langle E_{j}|E_{j_1}\rangle -\langle E_{j}|\Psi\rangle\langle\Psi|E_{j_1}\rangle
= 0
\]
due to $j_1\ne j$ and due to $\langle E_j|\Psi\rangle=0$.
Repeating this argument we find that the value $j$ drops out of all summations in Eq. (\ref{matrix_proda}),
reducing the accessible Hilbert space by 1 again.

With these arguments we have removed the degeneracy of the energy levels and the vanishing overlaps $p_j$. 
Therefore, the remaining return amplitude $\phi_k$ depends only non-degenerate energy levels and on overlaps with $p_j>0$.
For the subsequent analysis we use the convention that $N$ is the dimensionality of the accessible Hilbert space.


For the subsequent calculations it is useful to introduce 
two types of discrete Fourier transformations, where one is based on the phase factors $e^{i\omega k}$
\beq
{\tilde\phi}(\omega)=\sum_{k\ge 1} e^{i\omega k}\phi_k
\label{FT_u_phi}
\eeq
and the other is based on the random phase factors $e^{i\omega(\tau_1+\cdots+\tau_k)}\equiv e^{i\omega t_k}$
\beq
{\tilde\phi}_\tau(\omega)=\sum_{k\ge 1} e^{i\omega t_k}\phi_k
\ ,
\label{FT_u_phi2}
\eeq
provided that these series exist. Both ${\tilde\phi}(\omega)$ and ${\tilde\phi}_\tau(\omega)$ are still functions
of the random variables $\{\tau_k\}$.

\section{Summary and Results}
\label{sect:summary}

It will be shown that for a quantum system with energy levels $\{E_j\}$ and eigenstates $\{|E_j\rangle\}$ of 
a given Hamiltonian $H$
the FDR probability $|\phi_k|^2$ in the case of a ME is determined by the random phase factors 
$\{e^{-iE_j\tau_k}\}$ and the overlaps $\{|\langle E_j|\Psi\rangle|^2\}\equiv \{p_j\}$ alone. After averaging with
respect to the random measurements we get $\langle|\phi_k|^2\rangle_\tau$, which will turn out to be a function of
the $N$ parameters $\{\langle e^{-iE_j\tau}\rangle_\tau\}$ and of the $N^2$ parameters 
$\{\langle e^{-i(E_j-E_{j'})\tau}\rangle_\tau\}$. (Details are given in Sect. \ref{sect:matrix}.)
Using the Fourier transformations (\ref{FT_u_phi}) and (\ref{FT_u_phi2}), we define the generating functions 
$F(\omega)=\sum_{k\ge 1}e^{ik\omega}\langle|\phi_k|^2\rangle_\tau$ 
and $F_\tau(\omega)=\sum_{k\ge 1}\langle e^{i\omega t_k}|\phi_k|^2\rangle_\tau$. 
By differentation with respect to $\omega$ we get the mean number of measurements for the FDR and the mean 
FDR time as
\[
\overline{k}
=-i\partial_\omega F(\omega)|_{\omega=0}
\ ,\ \ 
\overline{t}=-i\partial_\omega F_\tau(\omega)|_{\omega=0}
\ .
\]
Then we will derive the relation $\overline{t}=\langle\tau\rangle_\tau\overline{k}$, where 
$\langle\tau\rangle_\tau$ is the mean time interval between successive measurements.

Besides the average FDR probability $\langle|\phi_k|^2\rangle_\tau$ we will also calculate the corresponding
expressions of the Fourier transform ${\tilde\phi}(\omega)$, namely $\langle|{\tilde\phi}(\omega)|^2\rangle_\tau$.
This will turn out to be 1, as shown in Eq. (\ref{norm1}), which is
essential for calculating the average Berry phase and the mean values of the FDR as
\beq
\overline{k}=N
\ , \ \ 
\overline{t}=\langle\tau\rangle_\tau N .
\label{exp01}
\eeq
The main result are listed in Table \ref{table1}.
%

\section{Fixed time step $\tau$}
\label{sect:fixed_time}

In this section we briefly recapitulate what is known about the FDR problem in the case of
non-random $\tau$ to set the stage for our investigation of random time steps $\{\tau_k\}$.
Stroboscopic measurement with $\tau_k=\tau$ has been studied extensively in the 
literature~\cite{Gruenbaum2013,Grunbaum2014,Friedman_2016,Friedman2017,thiel18,Yin2019}.
Next we summarize relevant information from previous works, in particular,
some results of Ref. \cite{Yin2019}.

At fixed $\tau$ the Laplace transformation for the return amplitude of the ME reads
\beq
{\hat\phi}(z)\equiv\sum_{k\ge1}z^k\phi_{k}
=\sum_{k\ge1}z^k\langle\Psi|(e^{-iH\tau}P)^{k-1}e^{-iH\tau}|\Psi\rangle
=\langle\Psi|(e^{i\tau H}/z-P)^{-1}|\Psi\rangle
\ .
\label{ret_amp}
\eeq
Due to the relation of Eq. (\ref{identity5}) in App. \ref{app:recursion},
we obtain for $K=e^{i\tau H}/z-{\bf 1}$ the following identity
\beq
\langle\Psi|(K+|\Psi\rangle\langle\Psi|)^{-1}|\Psi\rangle
=1-\frac{1}{1+\langle\Psi|K^{-1}|\Psi\rangle}
\ .
\label{identity6}
\eeq
This identity is important because it allows us to represent the projector-dependent left-hand side
by the expression $\langle\Psi|K^{-1}|\Psi\rangle$, which is diagonal in terms of the energy 
eigenstates of $H$ and independent of the projector $|\Psi\rangle\langle\Psi|$. A corresponding
identity exists in the case of random time steps, which will be central for our subsequent calculations.

With $u_k$ of Eq. (\ref{ue_amp}) we can write $\langle\Psi|K^{-1}|\Psi\rangle$
as the Laplace transform of $u_k$: $\langle\Psi|K^{-1}|\Psi\rangle=\sum_{k\ge 1}z^k u_k\equiv {\hat u}(z)$. 
Then ${\hat\phi}(z)$ in Eq. (\ref{ret_amp}), together with Eq. (\ref{identity6}), becomes
\beq
{\hat\phi}(z)=1-\frac{1}{1+{\hat u}(z)}
\ .
\eeq
By analytic continuation to the unit circle $z\to e^{i\omega}$ we get ${\hat u}(z)\to {\tilde u}(\omega)$
and ${\hat\phi}(z)\to {\tilde\phi}(\omega)$.
Since $Re[{\tilde u}(\omega)]=-1/2$, the expression ${\tilde \phi}$ is unimodular: 
\beq
{\tilde \phi}(\omega)=\frac{-1/2+i Im[{\tilde u}]}{1/2+iIm[{\tilde u}]}
=-\frac{{\tilde u}}{{\tilde u}^*}
\ ,
\label{unimodular_f}
\eeq
such that we can write
\beq
{\tilde \phi}(\omega)=-e^{2i \arg[{\tilde u}(\omega)]}\equiv e^{i\varphi(\omega)}
\ .
\label{phase_factor}
\eeq
This result for fixed time steps indicates that the UE and the ME have the same phase change with $\omega$ except for a 
factor 2. 
The winding number of ${\tilde \phi}(\omega)$ around the unit circle (i.e. for $0\le\omega<2\pi$) is identical with 
$\sum_{k\ge 1}k|\phi_k|^2$, and it is known that the winding number is equal to the dimensionality of the 
Hilbert space~\cite{Gruenbaum2013}.

\section{Matrix products}
\label{sect:matrix}

Our goal is to calculate the probability $|\phi_k|^2$ 
of the return amplitude $\phi_k$ of Eq. (\ref{matrix_proda}) for the general case of random time steps.
We would expect that the calculations of the previous section can be extended to this situation. As we will see though
it requires some additional steps to calculate quantities, such as the mean FDR time, that are averaged with respect to the 
random time steps.
Our calculation starts with the matrix representation of the projector $P$ of Eq. (\ref{me_amp}) in terms of energy 
eigenstates
\beq
\langle E_j|P|E_{j'}\rangle =
\delta_{j,j'}-q_{j} q_{j'}^*
\ ,\ \ 
q_{j}=\langle E_j|\Psi\rangle
\ ,
\eeq
since the eigenstates are orthonormal: $\langle E_j|E_{j'}\rangle= \delta_{j,j'}$.
This is automatically fulfilled for non-degenerate eigenvalues. 
The above expression is inserted in Eq. (\ref{matrix_proda}) and yields for 
the return amplitude a trace of a matrix product:
\[
\phi_k=Tr\left[
D_k({\bf 1}-QEQ^*)D_{k-1}({\bf 1}-QEQ^*)\cdots D_2({\bf 1}-QEQ^*)D_1QEQ^*\right]
\]
with the $N\times N$ matrix $E$, whose elements are all 1, and with the
diagonal matrices $D_k=diag(\exp(-iE_1\tau_k),\exp(-iE_2\tau_k),..., \exp(-iE_N\tau_k))$ 
and $Q=diag(q_{1},q_{2},..., q_{N})$. 

Now $QQ^*=\Pi $ is the diagonal matrix $\Pi =diag(p_{1},p_{2},...,p_{N})$, 
which enables us to rewrite $\phi_k$ as
\beq
\phi_k=Tr\left[D_k({\bf 1}-E\Pi )D_{k-1}({\bf 1}-E\Pi )\cdots D_2({\bf 1}-E\Pi )D_1E\Pi \right]
\label{matrix_prod2}
\ ,
\eeq
since $\Pi$ and $D_j$ as diagonal matrices commute. 
This means that $\phi_k$ depends only on the spectral weights $\{ p_j\}$ through $\Pi$ and on the energy 
levels $\{ E_j\}$ through $D_k$.

For the calculation of the return probability we need the product of two traces 
\[
|\phi_{k}|^2=Tr\left[D_k({\bf 1}-E\Pi )D_{k-1}({\bf 1}-E\Pi )\cdots D_2({\bf 1}-E\Pi )D_1E\Pi \right]^*
\]
\beq
\times
Tr\left[D_k({\bf 1}-E\Pi )D_{k-1}({\bf 1}-E\Pi )\cdots D_2({\bf 1}-E\Pi )D_1E\Pi \right]
\ .
\label{product_suma}
\eeq
In order to express this product it is convenient to use the notation of the Kronecker product of matrices
\[
{\hat A}=A_1\times A_2
\]
with the properties
\beq
{\hat A}{\hat B}=(A_1\times A_2)(B_1\times B_2)=A_1B_1\times A_2B_2
\ ,\ \
Tr({\hat A})=Tr(A_1)Tr(A_2)
\ , \ \
(A_1\times A_2)^{-1}=A_1^{-1}\times A_2^{-1} 
\ .
\label{cart_prop}
\eeq
The second identity, or trace ``disentanglement'' relation, is relevant for Eq. (\ref{product_suma}).
With the matrix
$
{\hat C}=({\bf 1}-E\Pi)\times({\bf 1}-E\Pi)
$
it gives us
\beq
|\phi_{k}|^2=Tr({\hat D}_k{\hat C}\cdots {\hat D}_2{\hat C}{\hat D}_1{\hat E}{\hat \Pi })
\label{product_sum_2}
\label{prob_distr}
\eeq
with ${\hat E}=E\times E$, ${\hat\Pi}=\Pi\times\Pi$ and ${\hat D}_k=D^*_{k}\times D_k$.
For the matrix elements we use the notation
\beq
[A\times B]_{ij,kl}=A_{ik}B_{jl}
\ .
\label{components}
\eeq

\section{Averaging over the distribution of random time steps}
\label{sect:time_av}

In the previous section we obtained a random distribution of return probabilities $\{|\phi_k|^2\}$.
Here we are interested in the mean values $\{\langle|\phi_k|^2\rangle_\tau\}$,
The subsequent calculation of the time average is based on the fact that the random 
matrices $\{{\hat D}_k\}$ are statistically independent and identically distributed. Thus, from
Eq. (\ref{product_sum_2}) we get
\beq
\langle|\phi_{k}|^2\rangle_\tau
=Tr(\langle{\hat D}_k\rangle_\tau{\hat C}\cdots \langle{\hat D}_2\rangle_\tau
{\hat C}\langle{\hat D}_1\rangle_\tau{\hat E}{\hat \Pi })
=Tr([\langle{\hat D}\rangle_\tau{\hat C}]^{k-1}\langle{\hat D}\rangle_\tau{\hat E}{\hat \Pi })
\ .
\label{phi^2}
\eeq
With the matrices
\beq
{\hat\Gamma}=\langle D^*({\bf 1}-E\Pi )\times D({\bf 1}-E\Pi )\rangle_\tau
=\langle D^*\times D\rangle_\tau ({\bf 1}-E\Pi )\times ({\bf 1}-E\Pi )
=\langle{\hat D}\rangle_\tau {\hat C}
\label{Gamma}
\eeq
and
\beq
{\hat G}=\langle D^*E\Pi \times DE\Pi \rangle_\tau
=\langle D^*\times D \rangle_\tau E\Pi\times E\Pi
=\langle{\hat D}\rangle_\tau{\hat E}{\hat \Pi}
\label{hG}
\eeq
we obtain the compact expression
\beq
\langle|\phi_{k}|^2\rangle_\tau
=Tr[{\hat\Gamma}^{k-1}{\hat G}]
\ .
\label{product_sum_3}
\eeq
${\hat\Gamma}$ depends on the averaged product $\langle D^*\times D\rangle_\tau$.
The latter cannot be expressed as a Kronecker product, which prevents us also from applying the trace 
''disentanglement'' relation of Eq. (\ref{cart_prop}). This reflects a robust ``entanglement'' due to 
$\langle D^*\times D\rangle_\tau$. We will return to this fact in the next section.

\subsection{The generating functions} 
\label{sect:gen_funct}

First, from Eq. (\ref{product_sum_3}) we obtain, after a discrete Fourier transformation, the generating functions
\beq
F(\omega)=\sum_{k\ge 1}e^{ik\omega}\langle|\phi_k|^2\rangle_\tau
=e^{i\omega}Tr[(\hat{\bf 1}-e^{i\omega}{\hat\Gamma})^{-1}{\hat G}]
\ ,
\label{generating_func00a}
\eeq
and
\beq
F_\tau(\omega)=\sum_{k\ge 1}\langle e^{i\omega(\tau_1+\cdots +\tau_k)}|\phi_k|^2\rangle_\tau
=Tr[(\hat{\bf 1}-{\hat\Gamma}_\omega)^{-1}{\hat G}_\omega]
\label{generating_func00b}
\eeq
with
\beq
{\hat\Gamma}_\omega
=\langle e^{i\omega\tau} {\hat D}\rangle_\tau{\hat C}
\eeq
and
\beq
{\hat G}_\omega
=\langle e^{i\omega\tau}{\hat D} \rangle_\tau E\Pi\times E\Pi
\ .
\eeq
As shown in App. \ref{app:gamma_an}, the matrix $(\hat{\bf 1}-z{\hat\Gamma})^{-1}$ is analytic for $|z|<1$.
This means that we should consider the discrete Fourier summation as an analytic continuation $z\to e^{i\omega}$
from $|z|<1$. 

From $F(\omega)$ and $F_\tau(\omega)$ we can calculate moments of $k$ and $t_k=\tau_1+\cdots +\tau_k$ with 
respect to the weight $\langle|\phi_k|^2\rangle_\tau$ and $\langle\tau|\phi_k|^2\rangle_\tau$, respectively, as 
\beq
\overline{k^m}=\sum_{k\ge 1}k^m\langle|\phi_k|^2\rangle_\tau=(-i\partial_\omega)^mF(\omega)|_{\omega=0}
\ ,\ \ 
\overline{t^m}=\sum_{k\ge 1}\langle t_k^m|\phi_k|^2\rangle_\tau=(-i\partial_\omega)^mF_\tau(\omega)|_{\omega=0}
\ .
\label{moments00}
\eeq
The property $F(\omega=0)=F_\tau(\omega=0)=1$, discussed in App. \ref{app:norm}, indicates a close relation between the
two generating functions. Then the $\omega$ dependence of the generating functions is through the fact that
(i) $F(\omega)$ and $F_\tau(\omega)$ depend on $\omega$ only through $\langle e^{i\omega}{\hat D}\rangle_\tau$ 
and $\langle e^{i\omega\tau}{\hat D}\rangle_\tau$, respectively,
and (ii) the matrix $\langle e^{i\omega\tau}{\hat D}\rangle_\tau$ is a function of $\omega+E_j-E_{j'}$.
It implies that we can replace a derivative with respect to $\omega$ by a derivative with respect to the difference of
energy levels if $E_j-E_{j'}\ne0$ ($j'\ne j$). Since the latter is implicitly assumed here for all energy levels, 
we can write for the first moment in Eq. (\ref{moments00})
\[
-i\partial_\omega F(\omega)\Big|_{\omega=0}=\sum_{j,j'}[\partial_{{\bar D}_{jj'}}F(\omega)]
\partial_\omega{\bar D}_{jj'}(\omega)\Big|_{\omega=0}
\]
\[
=\sum_{j\ne j'}[\partial_{{\bar D}_{jj'}(0)}F(0)]\partial_{E_j-E_{j'}}{\bar D}_{jj'}(0)
+\sum_j[\partial_{{\bar D}_{jj}(\omega)}F(\omega)]\partial_{\omega}{\bar D}_{jj}(\omega)\Big|_{\omega=0}
\]
with ${\bar D}_{jj'}(\omega)=e^{i\omega}\langle {\hat D}_{jj'}\rangle_\tau$.
The first sum on the right-hand side vanishes due to $F(0)=1$ and consequently $\partial_{E_j-E_{j'}} F(0)=0$,
such that we obtain
\beq
\partial_\omega F(\omega)\Big|_{\omega=0}=\sum_j\partial_{{\bar D}_{jj}(\omega)}F(\omega)\Big|_{\omega=0}
i\langle{\hat D}_{jj}(0)\rangle_\tau
=i\sum_j\partial_{{\bar D}_{jj}(\omega)}F(\omega)\Big|_{\omega=0}
\ .
\label{moment1}
\eeq
The analog calculation is valid for $F_\tau(\omega)$ and gives
\beq
\partial_\omega F_\tau(\omega)\Big|_{\omega=0}=\sum_j\partial_{{\bar D}_{jj}(\omega)}F_\tau(\omega)\Big|_{\omega=0}
i\langle\tau{\hat D}_{jj}(0)\rangle_\tau
=i\langle\tau\rangle_\tau\sum_j\partial_{{\bar D}_{jj}(\omega)}F(\omega)\Big|_{\omega=0}
\ .
\label{moment2}
\eeq
Comparing the expressions in Eqs. (\ref{moment1}) and (\ref{moment2}) implies
for the first moments in Eq. (\ref{moments00}) the relation
\beq
\overline{t}=
\sum_{k\ge 1}\langle t_k|\phi_k|^2\rangle_\tau=\langle\tau\rangle_\tau \sum_{k\ge 1}k\langle|\phi_k|^2\rangle_\tau
=\langle\tau\rangle_\tau\overline{k}
\ .
\label{mean_fdr}
\eeq


\subsection{Evaluation of $\langle|{\tilde\phi}(\omega)|^2\rangle_\tau$}
\label{sect:quantization1}

Next we will show that $\langle|{\tilde\phi}(\omega)|^2\rangle_\tau=1$ holds in general for any integer $N$ due to
\beq
T_{j_1j_2}
=\sum_{j_3,j_4}[\langle{\hat D}\rangle_\tau^{-1}
-{\hat C}]^{-1}_{j_1j_2,j_3j_4}
=\frac{1}{p_{j_1}}\delta_{j_1j_2}
\ ,
\label{sequence}
\eeq
where ${\hat C}$ was defined in Sect. \ref{sect:matrix}. 
To derive this property and to calculate the mean FDR time we use the matrix relations
\beq
{\hat E}{\hat W}[\langle{\hat D}\rangle_\tau^{-1}-{\hat C}]={\hat E}{\hat\Pi}
\ ,\ \ 
[\langle{\hat D}\rangle_\tau^{-1}-{\hat C}]{\hat T}{\hat E}={\hat E}
\ ,
\label{iden9}
\eeq
where ${\hat W}$ is the $N^2\times N^2$ diagonal matrix $diag(p_1,0_N,p_2,0_N,...,0_N,p_N)$, and
$0_N$ is a sequence of $N$ zeros. ${\hat T}$ is the $N^2\times N^2$ diagonal matrix with elements
$T_{jj'}=\delta_{jj'}/p_{j}$ of Eq. (\ref{sequence}). 
The $\langle{\hat D}\rangle_\tau$ contribution on the right-hand side disappears,
since ${\hat W}\langle{\hat D}\rangle_\tau={\hat W}$.
The first relation of Eq. (\ref{iden9}) is obtained from $W_{j_1j_2}=p_{j_1}\delta_{j_1j_2}$
by a direct inspection of the matrix elements:
\[
\sum_{j_1,j_2}W_{j_1j_2}[-E_{j_1j_3}p_{j_3}E_{j_2j_4}p_{j_4}+E_{j_1j_3}p_{j_3}\delta_{j_2j_4}
+\delta_{j_1j_3}E_{j_2j_4}p_{j_4}]=p_{j_3}p_{j_4}
\]
and the second relation for $T_{j_3j_4}=\delta_{j_3j_4}/p_{j_3}$ from
\[
\sum_{j_3,j_4}[-E_{j_1j_3}p_{j_3}E_{j_2j_4}p_{j_4}+E_{j_1j_3}p_{j_3}\delta_{j_2j_4}
+\delta_{j_1j_3}E_{j_2j_4}p_{j_4}]T_{j_3j_4}=1
\ .
\]
Provided that the inverse of $\langle{\hat D}\rangle_\tau^{-1}-{\hat C}$ exists,
the second relation of Eq. (\ref{iden9}) implies
$[\langle{\hat D}\rangle_\tau^{-1}-{\hat C}]^{-1}{\hat E}={\hat T}{\hat E}$, which
gives directly Eq. (\ref{sequence}) and subsequently the normalization
\beq
\langle|{\tilde\phi}(\omega)|^2\rangle_\tau=1
\label{norm1}
\eeq 
according to Eq. (\ref{norm2}) of App. \ref{app:gener_funct}.
%
This result can be used to reduce the integral of the average winding number
\beq
\langle w\rangle_\tau
:=\frac{1}{2\pi}\int_0^{2\pi}
\frac{\langle{\tilde \phi}(\omega)^*[-i\partial_\omega]{\tilde \phi}(\omega)\rangle_\tau}
{\langle|{\tilde \phi}(\omega)|^2\rangle_\tau}d\omega
\ ,
\label{berry_phase}
\eeq 
which is discussed in more detail in Sect. \ref{sect:discussion},
with the help of Eq. (\ref{norm1}) to
\beq
\langle w\rangle_\tau
=\frac{1}{2\pi}\int_0^{2\pi}
\langle{\tilde \phi}(\omega)^*[-i\partial_\omega]{\tilde \phi}(\omega)\rangle_\tau d\omega
=\sum_{k\ge 1}k\langle |\phi_k|^2\rangle_\tau=\overline{k}
\ .
\label{phi_1}
\eeq
In the next section we will see that $\langle w\rangle_\tau=\overline{k}$ is an integer, equal to the dimensionality of 
the accessible Hilbert space.
%

\subsubsection{Mean FDR time}
\label{sect:mean}

Now we return to the first moment in Eq. (\ref{moments00}), using an extension of the previous calculation.
Starting with
\beq
-i\partial_\omega F(\omega)\Big|_{\omega=0}
=Tr[(\hat{\bf 1}-{\hat\Gamma})^{-2}\langle{\hat D}\rangle_\tau{\hat E}{\hat\Pi}]
\label{moment3}
\eeq
we write for the matrix inside the trace
\[
{\hat E}{\hat\Pi}[\langle{\hat D}\rangle_\tau^{-1}-{\hat C}]^{-1}
\langle{\hat D}\rangle_\tau^{-1}[\langle{\hat D}\rangle_\tau^{-1}-{\hat C}]^{-1}
\]
and apply the first relation of Eq. (\ref{iden9}) to the first inverse matrix to obtain
\beq
={\hat E}{\hat W}\langle{\hat D}\rangle_\tau^{-1}
[\langle{\hat D}\rangle_\tau^{-1}-{\hat C}]^{-1}
={\hat E}{\hat W}[\langle{\hat D}\rangle_\tau^{-1}-{\hat C}]^{-1}
\ ,
\label{iden10}
\eeq
where the last equation is due to ${\hat W}\langle{\hat D}\rangle_\tau^{-1}={\hat W}$.
This can be inserted into Eq. (\ref{moment3}), and with the second relation of Eq. (\ref{iden9}) we get
\beq
-i\partial_\omega F(\omega)\Big|_{\omega=0}
=Tr\{{\hat E}{\hat W}[\langle{\hat D}\rangle_\tau^{-1}-{\hat C}]^{-1}\}
=Tr\{{\hat W}{\hat T}{\hat E}\}
=\sum_{j_1=1}^N T_{j_1j_1}p_{j_1}=N
\ ,
\label{wind_N}
\eeq
where the last two equations follow from Eq. (\ref{sequence}) and the definition of ${\hat W}$.
This result gives us, together with Eqs. (\ref{mean_fdr}) and (\ref{wind_N}),
for the mean number of measurements (MNM) of the FDR and the mean FDR time
\beq
\overline{k}=\sum_{k\ge 1}k\langle|\phi_k|^2\rangle_\tau =\langle w\rangle_\tau=N
\ \ {\rm and}\ \ \ 
\overline{t}=\sum_{k\ge 1}\langle t_k|\phi_k|^2\rangle_\tau=\langle\tau\rangle_\tau \langle w\rangle_\tau =\langle\tau\rangle_\tau N
\ ,
\label{mean-results}
\eeq
which presents an extension of a central result of the seminal work by Gr\"unbaum et al. \cite{Gruenbaum2013} 
of stroboscopic measurements to random time measurements.

As already mentioned in the Introduction, higher order moments are not quantized but can be very sensitive to degeneracies 
of the spectrum, at least for stroboscopic measurements\cite{Grunbaum2014,Friedman2017,Yin2019}. 
In the case of random measurements this is also true near degeneracies of the energy levels when 
$\langle D^*_j D_{j'}\rangle_\tau=\langle e^{i(E_j-E_{j'})\tau}\rangle_\tau$ ($j'\ne j$) is close to 1.
This originates in the fact that for
\[
-\partial_\omega^2 F(\omega)\Big|_{\omega=0}
=Tr[(\hat{\bf 1}-{\hat\Gamma})^{-3}\langle{\hat D}\rangle_\tau{\hat E}{\hat\Pi}]
=Tr\{{\hat W}(\hat{\bf 1}-{\hat\Gamma})^{-1}{\hat T}{\hat E}\}
\]
small eigenvalues of $\hat{\bf 1}-{\hat\Gamma}$ can appear.
This can indeed happen when at least one $\langle D^*_j D_{j'}\rangle_\tau$ is close to 1 or when one $p_j$ is
close to 0, as shown in App. \ref{app:gamma_an}.

\section{Example: symmetric two-level system}
\label{sect:example}

In the previous section we derived relations between the MNM of the FDR and the mean FDR time,
their relation with the average winding number $\langle w\rangle_\tau$ of Eq. (\ref{berry_phase})
and with the dimensionality of the Hilbert space  in Eq. (\ref{mean-results}).
These results are general and valid for any quantum system on an $N$--dimensional Hilbert space. Besides these mean
values we obtained in Eq. (\ref{moments00}) also higher moments of the number of measurements and the return time for the FDR.
For $m>1$ we have not found a simple expression but can obtain these moments only by calculating the
generating functions $F(\omega)$ and $F_\tau(\omega)$ directly. To determine these generating functions would require
the inversion of the $N^2\times N^2$ matrix $\hat{\bf 1}-z{\hat\Gamma}$. This is a tedious task, which goes beyond the scope 
of this paper. Therefore, we limit ourselves to $N=2$ and calculate the corresponding $4\times 4$ matrices 
(cf. App. \ref{app:2ls}). In particular,
we consider a symmetric two-level system (2LS) with energy levels $E_\pm=\pm J$ and spectral weights $p_1=p_2=1/2$
for random times. The Hilbert space is two-dimensional with two basis states, e.g. 
$|0\rangle$ and $|1\rangle$. If the measured state is $|0\rangle$, the projector $P$ reads $P=|1\rangle\langle 1|$.
Then we get $\langle 0|e^{-iH\tau}|0\rangle=\cos(J\tau)$ and $\langle 0|e^{-iH\tau}|1\rangle=i\sin(J\tau)$ and the
return amplitude $\phi_k$ reads
\beq
\phi_k=\cases{
\cos (J\tau_1) & for $k=1$ \cr
-\sin (J\tau_1)\sin(J\tau_2) & for $k=2$ \cr
-\sin (J\tau_1)\cos(J\tau_2)\cdots\cos(J\tau_{k-1})\sin(J\tau_k) & for $k\ge 3$ \cr
}
\ .
\label{2ls_ampl}
\eeq
The simplicity of the two-level system is manifested in the fact that $\phi_k$ is a scalar product
in contrast to the matrix product in Eq. (\ref{matrix_prod2}) of the general case $N>2$. This
simplifies calculations with respect to random $\{\tau_k\}$ substantially. For instance, we can easily
perform the summation with respect to $k$ to get 
\beq
\sum_{k=1}^n|\phi_{k}|^2=1-[1-\cos^2 (J\tau_1)]\cos^2(J\tau_2)\cdots\cos^2(J\tau_{n})
\label{norm_2l}
\eeq
as a special case of the general equation (\ref{norm}).
After averaging with respect to $\{\tau_k\}$, we get for the mean FDR time
\beq
\overline{t}
=\sum_{k\ge 1}\langle(\tau_1+\cdots +\tau_k)|\phi_k|^2\rangle_\tau =2\langle\tau\rangle_\tau
\ ,
\eeq
which is in agreement with our general result in Eq. (\ref{mean-results}).
Moreover, the generating functions $F(\omega)$ in Eq. (\ref{generating_func00a}) and
$F_\tau(\omega)$ in Eq. (\ref{generating_func00b}) read
\beq
F(\omega)
=e^{i\omega}\frac{(2e^{i\omega}-1)\langle\cos2J\tau\rangle_\tau-1}{e^{i\omega}(\langle\cos2J\tau\rangle_\tau+1)-2}
\ ,\ \ 
F_\tau(\omega)=\frac{(2\langle e^{i\omega\tau}\rangle_\tau-1)\langle e^{i\omega\tau}\cos2J\tau
\rangle_\tau-\langle e^{i\omega\tau}\rangle_\tau}
{\langle e^{i\omega\tau}\cos2J\tau\rangle_\tau+\langle e^{i\omega\tau}\rangle_\tau-2}
\ ,
\eeq
which gives for $\omega=0$
\beq
\sum_{k\ge 1}\langle|\phi_k|^2\rangle_\tau=F(0)=1
\ ,\ \
\sum_{k\ge 1}k\langle|\phi_k|^2\rangle_\tau=-iF'(0)=2
\ ,\ \ 
\sum_{k\ge 1}k^2\langle|\phi_k|^2\rangle_\tau=-F''(0)=2\frac{3-\langle\cos2J\tau\rangle_\tau}{1-\langle\cos2J\tau\rangle_\tau}
\ ,
\label{fluct}
\eeq
where the first equation reflects the normalization, the second equation the quantization of the mean FDR time and the
last equation the FDR fluctuations. The latter only diverge for $J=0$ in the case of random times steps, but
in the case of a fixed time step $\tau$ it also diverges for $J\tau=k\pi$ ($k=\pm 1,\pm 2, ...$) . For $F_\tau(\omega)$ we get
\beq
\sum_{k\ge 1}\langle(\tau_1+\cdots+\tau_k)|\phi_k|^2\rangle_\tau=
-iF_\tau'(0)=2\langle\tau\rangle_\tau
\ .
\label{time}
\eeq

\begin{figure}[t]
\begin{center}
\includegraphics[width=7cm,height=7cm]{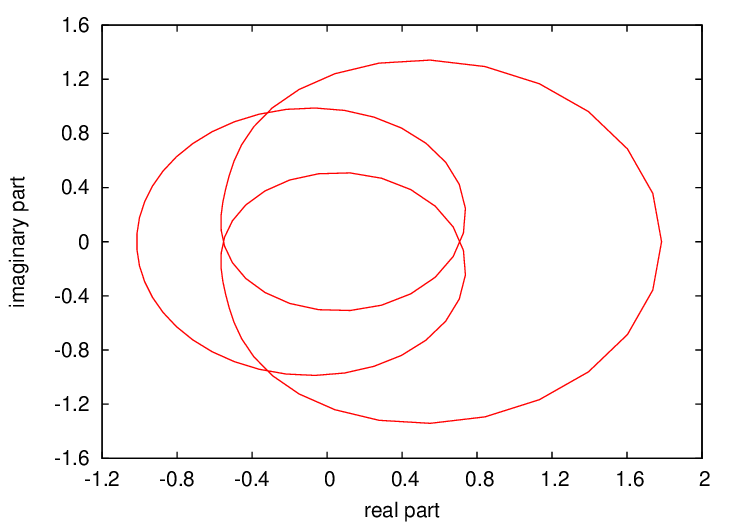}
\includegraphics[width=7cm,height=7cm]{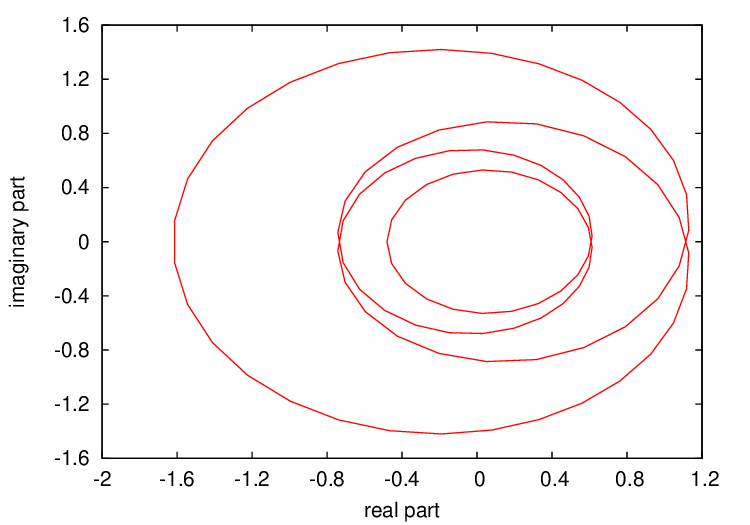}
\includegraphics[width=7cm,height=7cm]{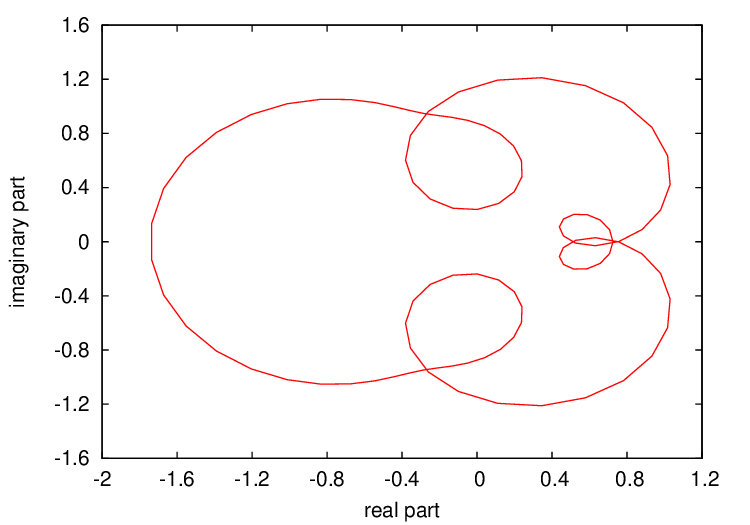}
\caption{
Symmetric two-level system:
The return amplitude ${\tilde\phi}_8(\omega)=\sum_{k=1}^8e^{ik\omega}\phi_k$ 
on the $\omega$ interval $[0,2\pi)$ 
with three randomly chosen realizations of $\{\tau_1,...,\tau_8\}$ performs a closed trajectory in the
complex plane. The winding numbers in these examples are $w_\phi=3,4,1$, respectively.
}
\label{fig:random}
\end{center}
\end{figure}

\begin{figure}[t]
\begin{center}
\includegraphics[width=9cm,height=7cm]{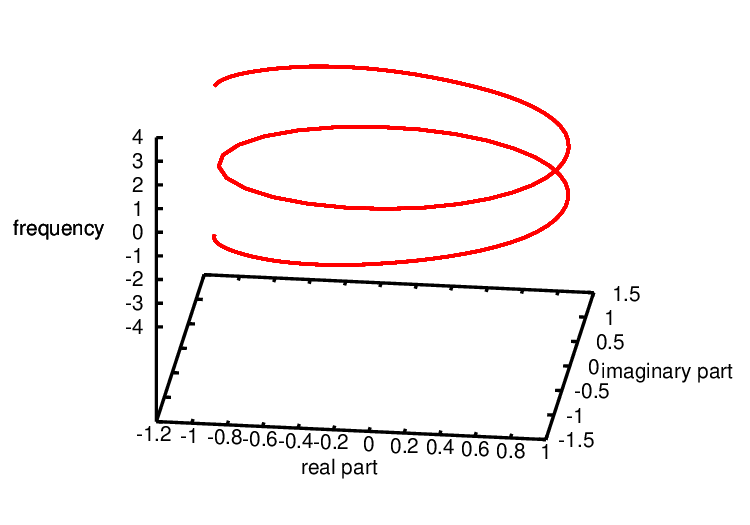}
\caption{
Symmetric two-level system:
The return amplitude ${\tilde\phi}_8(\omega)=\sum_{k=1}^8e^{ik\omega}\phi_k$ 
on the $\omega$ interval $[0,2\pi)$ with fixed time step $\tau\approx\pi/2J$. 
The winding number in this example is $w_\phi=2$.
}
\label{fig:stroboscopic}
\end{center}
\end{figure}

The two-level systems gives also a direct insight into the effect of random time steps
on the return amplitude ${\tilde\phi}(\omega)$ before averaging, since we can calculate these amplitudes 
from Eq. (\ref{2ls_ampl}) for special realizations of $\{\tau_k\}$. A few examples are visualized in Fig. 
\ref{fig:random}, indicating that $|{\tilde\phi}(\omega)|$ as well as the winding number vary from realization
to realization substantially. 
For fixed time steps, on the other hand, we have $|{\tilde\phi}(\omega)|=1$ and the
winding number is 2 in Fig. \ref{fig:stroboscopic}, as predicted by the general theory of stroboscopic 
measurements~\cite{Gruenbaum2013,Yin2019}.

\begin{figure}[t]
\begin{center}
\includegraphics[width=7cm,height=7cm]{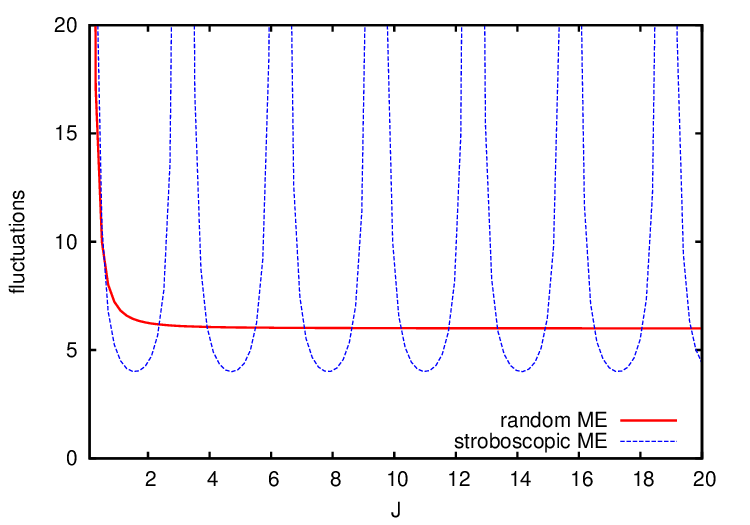}
\caption{
Symmetric two-level system: Fluctuations of the FDR for stroboscopic ME and random time steps
from Eqs. (\ref{rand_fluc}) and (\ref{strob_fluc}).
}
\label{fig:fluctuations}
\end{center}
\end{figure}

\section{Discussion}
\label{sect:discussion}

The aim of our work has been to calculate the properties of the ME with random measurements through the return 
amplitude ${\tilde \phi}(\omega)$. 
For given eigenvalues $\{E_j\}$ and eigenstates $\{|E_j\rangle\}$ of the Hamiltonian $H$
the ME of the return to the initial state $|\Psi\rangle$ are characterized by the time averaged phase factors
$\langle e^{\pm iE_j\tau}\rangle_\tau$, $\langle e^{ i(E_j-E_j')\tau}\rangle_\tau$ and by the spectral 
weights $p_j=|\langle E_j|\Psi\rangle|^2$. 
This allowed us to calculated the mean FDR times and the MNM of the FDR, using
the generating functions defined in Eqs. (\ref{generating_func00a}) and (\ref{generating_func00b}).
The surprising result in Eq. (\ref{mean-results}) is that the MNM of the FDR is just the dimensionality $N$ 
of the accessible 
Hilbert space and that the mean FDR time is $\langle\tau\rangle_\tau N$, where $\langle\tau\rangle_\tau$
is the mean time between two successive measurements. 
The same was previously observed for fixed time steps $\tau$~\cite{Gruenbaum2013,Grunbaum2014,Yin2019}. 
The robustness of the average winding number is remarkable, since the winding number of the return amplitude 
${\tilde\phi}(\omega)$ fluctuates strongly from realization to realization of the random $\{\tau_k\}$ (cf. Fig. \ref{fig:random}).

Other quantities, such as the correlation function $\langle \phi_k^*\phi_{k'}\rangle_\tau$ 
of the return amplitudes for different discrete times $k$ and $k'$ in Eq. (\ref{off-diag_phia}), can also be calculated.
An example is $\sum_{k,k'\ge1}e^{i(k'-k)\omega}\langle \phi_k^*\phi_{k'}\rangle_\tau
=\langle|{\tilde\phi}(\omega)|^2\rangle_\tau$, which is 1 
according to the calculation in App. \ref{app:gener_funct}. This enabled us to determine the mean FDR time
in Eq. (\ref{mean-results}). 
%
Its connection with the integral in Eq. (\ref{berry_phase}) is a generalization of the quantized
winding number $w_{st}$ in the case of stroboscopic measurements by Gr\"unbaum et al. \cite{Gruenbaum2013,Yin2019}. 
The latter is based on the fact that $|{\tilde\phi}(\omega)|=1$ (cf. Eq.(\ref{phase_factor})). Then the winding number $w_{st}$ 
simply reads for ${\tilde\phi}(\omega)=e^{i\varphi}$ (cf. Ref. \cite{Friedman2017} and Sect. \ref{sect:fixed_time})
\[
w_{st}=-\frac{i}{2\pi}\int_0^{2\pi}{\tilde\phi}(\omega)^*\partial_\omega{\tilde\phi}(\omega)d\omega
=\frac{1}{2\pi}\int_0^{2\pi}\partial_\omega\varphi d\omega
\ .
\]
In the case of random time measurements $|{\tilde\phi}(\omega;\{\tau_k\})|\ne1$, such that we must modify the definition
of the winding number by normalizing ${\tilde\phi}(\omega,\{\tau_k\})$. 
A further problem is that the winding number depends on the realization of the time steps $\{\tau_k\}$, as we have 
demonstrated in Fig. \ref{fig:random} for the symmetric two-level system. Therefore, we must also perform an average
with respect to the time steps. Then the definition of the winding number becomes Eq. (\ref{berry_phase}),
which reads with ${\tilde\phi}(\omega)=|{\tilde\phi}(\omega)|e^{i\varphi}$ as an average differential phase
change
\beq
\langle w\rangle_\tau
=\frac{1}{2\pi}\int_0^{2\pi}
\frac{\langle{|\tilde \phi}(\omega)|^2\partial_\omega\varphi\rangle_\tau}{\langle|{\tilde \phi}(\omega)|^2\rangle_\tau}d\omega
\ .
\eeq
Here we note that Eq. (\ref{berry_phase}) is formally equivalent to the definition of the Berry phase \cite{berry84} 
when we replace ${\tilde\phi}(\omega,\{\tau_k\})$ by the spatial wave function $\phi(\omega,\br)$ and replace the
time average $\langle ... \rangle_\tau$ by the usual quantum average in space. 

Averaging over random time steps is crucial to obtain a generic winding number. 
That different special realizations of the random time steps lead to different winding numbers can be seen
when we assume a finite sequence of $\{\phi_k\}$ ($k\le M$). The latter is either the result of an 
approximative truncation of the sequence or when the sequence terminates with $S_M=0$ in Eq. (\ref{norm}). 
Then the Fourier transformed return amplitude in Eq. (\ref{FT_u_phi}) becomes a finite sum 
\[
{\tilde\phi}_M(\omega)=\sum_{k=1}^M e^{i\omega k}\phi_k 
\]
for which a winding number $w_M$ can be defined for (random) coefficients $\{\phi_k\}$ as
\beq
w_M=\frac{1}{2\pi}\int_0^{2\pi}\frac{{\tilde\phi}^*_M(\omega)(-i\partial_\omega){\tilde\phi}_M(\omega)}
{{\tilde\phi}^*_M(\omega){\tilde\phi}_M(\omega)}d\omega
=\frac{1}{2\pi}\int_0^{2\pi}(-i\partial_\omega)\log[{\tilde\phi}_M(\omega)]d\omega
\ .
\eeq
Then we rewrite the polynomial ${\tilde\phi}_M(\omega)$ as the product
\[
{\tilde\phi}_M(\omega)=\phi_M(e^{i\omega}-z_1)\cdots(e^{i\omega}-z_M)
\ ,
\]
such that we get for the winding number
\[
w_M=\sum_{k=1}^M\frac{1}{2\pi}\int_0^{2\pi}\frac{e^{i\omega}}{e^{i\omega}-z_k}d\omega
\ .
\]
With $z=e^{i\omega}$ this gives
\[
w_M=\sum_{k=1}^M\frac{1}{2\pi i}\int_{S_1}\frac{1}{z-z_k}dz=M'
\ ,
\]
where the Cauchy integral is performed over the unit circle $S_1$ and $M'$ ($0\le M'\le M$) is the number of 
poles inside the unit circle. A simple example is $M=2$ with
\[
{\tilde\phi}_2(\omega)=\phi_1e^{i\omega}+\phi_2 e^{2i\omega}=z(\phi_1+\phi_2z)=\phi_2z(z+\phi_1/\phi_2)
\ ,
\]
where we get
\[
w_2=\cases{
1 & $|\phi_1/\phi_2|>1$ \cr
2 & $|\phi_1/\phi_2|<1$ \cr
}
\ .
\]
A detailed calculation of several quantities was presented in Sect. \ref{sect:example} for
the case of a symmetric two-level system with $w=2$ after averaging with respect to $\{\tau_k\}$. 
The fluctuations of the return time are finite
\beq
\sum_{k\ge 1}k^2\langle|\phi_k|^2\rangle_\tau
=2\frac{3-\langle\cos2J\tau\rangle_\tau}{1-\langle\cos2J\tau\rangle_\tau}
=\frac{1+6J^2}{J^2}
\ ,
\label{rand_fluc}
\eeq
where the last expression is obtained from the Poisson distribution $e^{-\tau_k}d\tau_k$. 
In the limit of a fixed measurement time $\tau$ the fluctuations 
\beq
\sum_{k\ge 1}k^2|\phi_k|^2=2\frac{3-\cos2J\tau}{1-\cos2J\tau}
\label{strob_fluc}
\eeq
would diverge for $J\tau=\pi n$ ($n=0,1,...$).
Thus, the random measurements wash out the divergences of the fluctuations. For most values of the level splitting
$J$ the fluctuations are stronger for the fixed time steps, as visualized in Fig. \ref{fig:fluctuations}.

In this paper we have completely focused on the return of the quantum system to its initial state. A natural
extension would be a corresponding analysis of the transition from an initial to a different final state,
monitored by random projective measurements. We have addressed this topic in a separate article~\cite{PhysRevA.103.022222}.

In conclusion, the mean FDR time of the ME for random time steps is equal to the dimensionality of the accessible 
Hilbert space. This is very similar to the ME for fixed time steps.
On the other hand, the strong fluctuations of the FDR time, which appear for a small distance of eigenvalues
in the case of fixed time steps, are washed out by averaging with respect to the random time steps. This was
briefly discussed for a two-level system in this article and more general in Ref. \cite{PhysRevA.103.022222}.
\begin{table}
\begin{tabular}{| l | l | l |}
\hline
 & \textbf{SME} & \textbf{RME}\\ \hline
probability $|{\tilde\phi}(\omega)|^2$ & $1$ & random  \\
probability $\langle|{\tilde\phi}(\omega)|^2\rangle_\tau$ & $-$ & $1$ \\
winding number $w$ of ${\tilde\phi}(\omega)$ & $N$ & random \\
winding number $\langle w\rangle_\tau$ of ${\tilde\phi}(\omega)$ & $-$ & $N$ \\ 
mean number of measurements for FDR $\overline{k}$ & $N$ & $N$ \\
mean FDR time $\overline{t}$ & $N\tau$ & $N\langle\tau\rangle_\tau$ \\
\hline
\end{tabular}
\caption{Comparison between stroboscopic ME (SME) and random ME (RME) with the Hilbert space dimensionality $N$.
The first detected return agrees for both approaches, provided that we average over the random measurements.}
\label{table1}
\end{table}

\vskip0.1cm
\no

\section*{Acknowledgments:} 
The support of Israel Science Foundation’s Grant No. 1898/17 as well as the support by the Julian
Schwinger Foundation (K.Z.) are acknowledged.

\appendix

\section{Normalization}
\label{app:norm}

The normalization of the vector ${\vec \phi}=(\phi_1,\phi_2,...,\phi_n)$ with
\[
\phi_{k}=\langle\Psi|e^{-iH\tau_k}(Pe^{-iH\tau_{k-1}})\cdots(Pe^{-iH\tau_1})|\Psi\rangle
\] 
in the limit $n\to\infty$ is based on the normalization of $|\Psi\rangle$
\[
\langle\Psi|e^{iH\tau_1}e^{-iH\tau_1}|\Psi\rangle=\langle\Psi|\Psi\rangle=1
\]
and will be derived by iteration:
For the evaluation of $|{\vec \phi}|^2$ we consider the sequence of projection operators $\{\Pi_k\}_{k=1,2,...,n}$ with
\[
\Pi_k := Pe^{iH\tau_k}e^{-iH\tau_k}P
\ ,\ \ 
P={\bf 1}-|\Psi\rangle\langle\Psi|\equiv {\bf 1}-P_0
\ .
\]
With $\Pi_k=P^2=P$ and $|\Psi\rangle\langle\Psi|+P={\bf 1}$ we can insert $P_0+\Pi_2={\bf 1}$ at
\[
1=\langle\Psi|e^{iH\tau_1}e^{-iH\tau_1}|\Psi\rangle
=\langle\Psi|e^{iH\tau_1}(P_0+\Pi_2)e^{-iH\tau_1}|\Psi\rangle
\]
\[
=\langle\Psi|e^{iH\tau_1}|\Psi\rangle\langle\Psi|e^{-iH\tau_1}|\Psi\rangle
+\langle\Psi|e^{iH\tau_1}\Pi_2e^{-iH\tau_1}|\Psi\rangle
\ .
\]
Next we replace $\Pi_2$ in the second term by 
\[
\Pi_2=Pe^{iH\tau_2}e^{-iH\tau_2}P=Pe^{iH\tau_2}(P_0+\Pi_3)e^{-iH\tau_2}P
\]
to get
\[
1=\langle\Psi|e^{iH\tau_1}|\Psi\rangle\langle\Psi|e^{-iH\tau_1}|\Psi\rangle
+\langle\Psi|e^{iH\tau_1}Pe^{iH\tau_2}(P_0+\Pi_3)e^{-iH\tau_2}Pe^{-iH\tau_1}|\Psi\rangle
\]
\[
=\langle\Psi|e^{iH\tau_1}|\Psi\rangle\langle\Psi|e^{-iH\tau_1}|\Psi\rangle
+\langle\Psi|e^{iH\tau_1}Pe^{iH\tau_2}|\Psi\rangle\langle\Psi|e^{-iH\tau_2}Pe^{-iH\tau_1}|\Psi\rangle
\]
\[
+\langle\Psi|e^{iH\tau_1}Pe^{iH\tau_2}\Pi_3e^{-iH\tau_2}Pe^{-iH\tau_1}|\Psi\rangle
\]
The replacement of the operator $\Pi_k$ by $Pe^{iH\tau_k}(P_0+\Pi_{k+1})e^{-iH\tau_k}P$
can be repeated for $k=3,...,n$ to obtain
\[
1=\langle\Psi|e^{iH\tau_1}|\Psi\rangle\langle\Psi|e^{-iH\tau_1}|\Psi\rangle
+\langle\Psi|e^{iH\tau_1}Pe^{iH\tau_2}|\Psi\rangle\langle\Psi|e^{-iH\tau_2}Pe^{-iH\tau_1}|\Psi\rangle
\]
\[
+\cdots+\langle\Psi|e^{iH\tau_1}(Pe^{iH\tau_2})\cdots(Pe^{iH\tau_n})|\Psi\rangle
\langle\Psi|(e^{-iH\tau_n}P)\cdots(e^{-iH\tau_2}P)e^{-iH\tau_1}|\Psi\rangle
\]
\[
+\langle\Psi|e^{iH\tau_1}(Pe^{iH\tau_2})\cdots(Pe^{iH\tau_n})P(e^{-iH\tau_n}P)\cdots(e^{-iH\tau_2}P)e^{-iH\tau_1}|\Psi\rangle
\]
\[
=\sum_{k=1}^n|\phi_k|^2+S_n
\ ,
\]
where
\[
S_n=\langle\Psi|e^{iH\tau_1}(Pe^{iH\tau_2})\cdots(Pe^{iH\tau_n})P(e^{-iH\tau_n}P)\cdots(e^{-iH\tau_2}P)e^{-iH\tau_1}|\Psi\rangle
\ .
\]
Thus, we have
\beq
\sum_{k=1}^n|\phi_{k}|^2=1-S_n
\ ,
\label{norm}
\eeq
where the probability $S_n$ of not recording the state after $n$ attempts is
\[
S_n=\langle\Psi|e^{iH\tau_1}(Pe^{iH\tau_2})\cdots(Pe^{iH\tau_n})P(e^{-iH\tau_n}P)
\cdots(e^{-iH\tau_2}P)e^{-iH\tau_1}|\Psi\rangle
\ .
\]
Provided the remainder $S_n$ vanishes in the limit $n\to\infty$, the wave function
${\vec\phi}=(\phi_1,\phi_2,...,\phi_n)$ is normalized in this limit: $|{\vec \phi}|^2=1$.
Although we do not have proof that $S_n$ always vanishes with $n\to\infty$, the latter is plausible 
and is supported by the example of the symmetric two-level system in Eq. (\ref{norm_2l}), 
by stroboscopic measurements \cite{Gruenbaum2013} and in the case of the time averaged sum 
$\sum_{k\ge 1}\langle|\phi_k|^2\rangle_\tau=1$ \cite{PhysRevA.103.022222}.

\section{Recursion}
\label{app:recursion}

For an operator $K$ and the projector $|\Psi\rangle\langle\Psi|$ we assume that the 
inverse $(K+|\Psi\rangle\langle\Psi|)^{-1}$ exists. Then we get the relation
\beq
\langle\Psi|(K+|\Psi\rangle\langle\Psi|)^{-1}|\Psi_0\rangle
=\langle\Psi|K^{-1}({\bf 1}+|\Psi\rangle\langle\Psi|K^{-1})^{-1}|\Psi_0\rangle
\ .
\label{rel1}
\eeq
Now we can use the identity
\[
({\bf 1}+|\Psi\rangle\langle\Psi|K^{-1})^{-1}={\bf 1}-|\Psi\rangle\langle\Psi|K^{-1}
({\bf 1}+|\Psi\rangle\langle\Psi|K^{-1})^{-1}
\]
to write for the right-hand side of Eq. (\ref{rel1})
\[
\langle\Psi|K^{-1}|\Psi_0\rangle-\langle\Psi|K^{-1}|\Psi\rangle
\langle\Psi|K^{-1}({\bf 1}+|\Psi\rangle\langle\Psi|K^{-1})^{-1}|\Psi_0\rangle
\ ,
\]
such that Eq. (\ref{rel1}) becomes
\[
\langle\Psi|(K+|\Psi\rangle\langle\Psi|)^{-1}|\Psi_0\rangle
=\langle\Psi|K^{-1}|\Psi_0\rangle-\langle\Psi|K^{-1}|\Psi\rangle
\langle\Psi|(K+|\Psi\rangle\langle\Psi|)^{-1}|\Psi_0\rangle
\ .
\]
This relation is equivalent to
\beq
\langle\Psi|(K+|\Psi\rangle\langle\Psi|)^{-1}|\Psi_0\rangle
=\frac{\langle\Psi|K^{-1}|\Psi_0\rangle}
{1+\langle\Psi|K^{-1}|\Psi\rangle}
\ .
\label{identity7}
\eeq
In particular, for the diagonal case $|\Psi_0\rangle=|\Psi\rangle$ we have
\beq
\langle\Psi|(K+|\Psi\rangle\langle\Psi|)^{-1}|\Psi\rangle=1-\frac{1}{1+\langle\Psi|K^{-1}|\Psi\rangle}
\ .
\label{identity5}
\eeq
These relations read in the energy (spectral) representation, where we assume that $K$ is diagonal,
\[
\langle\Psi|(K+|\Psi\rangle\langle\Psi|)^{-1}|\Psi_0\rangle
=\sum_{j,j'}\langle\Psi|E_j\rangle\langle E_j|(K+|\Psi\rangle\langle\Psi|)^{-1}|E_{j'}\rangle\langle E_{j'}|\Psi_0\rangle
=\sum_{j,j'}(K+\Pi E)^{-1}_{E_j,E_{j'}}\langle\Psi|E_{j'}\rangle\langle E_{j'}|\Psi_0\rangle
\]
and
\[
\langle\Psi|(K+|\Psi\rangle\langle\Psi|)^{-1}|\Psi\rangle
=\sum_{j,j'}\langle\Psi|E_j\rangle\langle E_j|(K+|\Psi\rangle\langle\Psi|)^{-1}|E_{j'}\rangle\langle E_{j'}|\Psi\rangle
=\sum_{j,j'}(K+\Pi E)^{-1}_{E_j,E_{j'}}\Pi_{E_{j'}}
\ ,
\]
where $K$ on the right-hand side is in the energy representation 
$\langle E_j|K|E_{j'}\rangle=K_{E_j,E_j}\delta_{E_j,E_{j'}}$.
This implies with Eq. (\ref{identity5})
\beq
\sum_{j,j'}(K+\Pi E)^{-1}_{E_j,E_{j'}}\Pi_{E_{j'}}
=1-\frac{1}{1+\sum_{j}K^{-1}_{E_j,E_j}\Pi_{E_{j}}}
\label{identity5a}
\eeq
and with Eq. (\ref{identity7})
\beq
\sum_{j,j'}(K+\Pi E)^{-1}_{E_j,E_{j'}}\langle\Psi|E_{j'}\rangle\langle E_{j'}|\Psi_0\rangle
=\frac{\sum_{j}K^{-1}_{E_j,E_{j}}\langle\Psi|E_{j}\rangle\langle E_{j}|\Psi_0\rangle}{1+\sum_{j}K^{-1}_{E_j,E_{j}}\Pi_{E_{j}}}
\ .
\label{identity7a}
\eeq
Here $|\Psi_0\rangle$ can be any state, implying that the relation holds for any
$Q_{E_{j'}}=\langle\Psi|E_{j'}\rangle h_{E_{j'}}\langle E_{j'}|\Psi\rangle =p_{j'}h_{E_{j'}}$:
\beq
\sum_{j,j'}(K+\Pi E)^{-1}_{E_j,E_{j'}}Q_{E_{j'}}
=\frac{\sum_{j}K^{-1}_{E_j,E_{j}}Q_{E_{j}}}{1+\sum_{j}K^{-1}_{E_j,E_{j}}\Pi_{E_{j}}}
\ .
\label{identity7b}
\eeq

\section{Product matrices}
\label{app:products2}

With the properties of Eq. (\ref{cart_prop}) we get for $k'\ge k$
\[
\phi^*_k\phi_{k'}=Tr[D^*_{k}({\bf 1}-E\Pi )\cdots D^*_2({\bf 1}-E\Pi )D_1^*E\Pi]
Tr[D_{k'}({\bf 1}-E\Pi )\cdots D_2({\bf 1}-E\Pi )D_1E\Pi]
\]
due to $Tr(A_1)Tr(A_2)=Tr(A_1\times A_2)$
\[
\phi^*_k\phi_{k'}=Tr\{[D^*_{k}({\bf 1}-E\Pi )\cdots D^*_2({\bf 1}-E\Pi )D_1^*E\Pi]
\times [D_{k'}({\bf 1}-E\Pi )\cdots D_2({\bf 1}-E\Pi )D_1E\Pi]\}
\]
and due to $A_1B_1\times A_2B_2=[A_1\times A_2][B_1\times B_2]$
\[
\phi^*_k\phi_{k'}=Tr\{\left[{\bf 1}\times D_{k'}({\bf 1}-E\Pi )\cdots {\bf 1}\times D_{k+1}({\bf 1}-E\Pi )\right]
\]
\[
\left[D^*_{k}({\bf 1}-E\Pi )\times D_{k}({\bf 1}-E\Pi )\cdots D^*_2({\bf 1}-E\Pi)\times D_2({\bf 1}-E\Pi)\right]
D_1^*E\Pi \times D_1 E\Pi\}
\ .
\]
Averaging with respect to independent random times $\{\tau_k\}$ then gives
\[
\langle\phi^*_k\phi_{k'}\rangle_\tau =Tr\{{\hat C}_2^{k'-k}{\hat\Gamma}^{k-1}{\hat G}\}
\ \ {\rm with}\ \ {\hat C}_2={\bf 1}\times \langle D\rangle_\tau({\bf 1}-E\Pi )
\ ,
\]
since 
\[
\langle D^*({\bf 1}-E\Pi )\times D({\bf 1}-E\Pi )\rangle_\tau
=\langle D^*\times D\rangle_\tau({\bf 1}-E\Pi )\times({\bf 1}-E\Pi )
\]
from the first relation in Eq. (\ref{cart_prop}) implies
\[
Tr\{{\bf 1}\times\langle D\rangle_\tau({\bf 1}-E\Pi )\cdots{\bf 1}\times \langle D\rangle_\tau({\bf 1}-E\Pi )
\langle D^*({\bf 1}-E\Pi )\times D({\bf 1}-E\Pi )\rangle_\tau\cdots\langle D^*({\bf 1}-E\Pi )\times D({\bf 1}-E\Pi)\rangle_\tau
\]
\[
\langle D^*E\Pi \times D E\Pi\rangle_\tau\}
\]
\[
=Tr\{{\bf 1}\times\langle D\rangle_\tau({\bf 1}-E\Pi )\cdots{\bf 1}\times \langle D\rangle_\tau({\bf 1}-E\Pi )
\langle D^*\times D\rangle_\tau({\bf 1}-E\Pi )\times({\bf 1}-E\Pi )\cdots\langle D^*\times D\rangle_\tau ({\bf 1}-E\Pi )\times({\bf 1}-E\Pi)
\]
\[
\langle D^*\times D\rangle_\tau E\Pi \times E\Pi\}
=Tr(
{\hat C}_2^{k'-k}{\hat\Gamma}^{k-1}{\hat G})
\ ,
\]
where the last equation follows from Eqs. (\ref{Gamma}), (\ref{hG}).
An analog expression exists for $k\ge k'$, such that we get
\beq
\langle \phi_k^*\phi_{k'}\rangle_\tau
=\cases{
Tr[({\hat C}_1)^{k-k'}{\hat\Gamma}^{k'-1}{\hat G}] & for $k\ge k'\ge 1$ \cr
Tr[({\hat C}_2)^{k'-k}{\hat\Gamma}^{k-1}{\hat G}] & for $k'\ge k\ge 1$ \cr
}
\ ,\ \ 
{\hat C}_j=\cases{
\langle D^*\rangle_\tau({\bf 1}-E\Pi )\times{\bf 1} & $j=1$ \cr
 {\bf 1}\times \langle D\rangle_\tau({\bf 1}-E\Pi ) & $j=2$ \cr
}
\ .
\label{off-diag_phia}
\eeq

\section{Generating function}
\label{app:gener_funct}

Next we consider the Fourier transform of Eq. (\ref{off-diag_phia})
\beq
\langle {\tilde\phi}^*(\omega){\tilde\phi}(\omega+\omega')\rangle_\tau
=\sum_{k\ge 1}\sum_{k'\ge k}e^{i(\omega+\omega')(k'-k)+i\omega'k}\langle \phi_k^*\phi_{k'}\rangle_\tau
+\sum_{k'\ge 1}\sum_{k>k'}e^{i\omega(k'-k)+i\omega'k'}\langle \phi_k^*\phi_{k'}\rangle_\tau
\label{om-corr}
\eeq
\[
=e^{i\omega'}
Tr\{[(\hat{\bf 1}-e^{i(\omega+\omega')}{\hat C}_2)^{-1}+e^{-i\omega}{\hat C}_1(\hat{\bf 1}-e^{-i\omega}{\hat C}_1)^{-1}]
(\hat{\bf 1}-e^{i\omega'}{\hat\Gamma})^{-1}{\hat G}\}
\ ,
\]
which becomes, after rewriting the second term in the trace,
\[
=e^{i\omega'}
Tr\{[(\hat{\bf 1}-e^{i(\omega+\omega')}{\hat C}_2)^{-1}+(\hat{\bf 1}-e^{-i\omega}{\hat C}_1)^{-1}-\hat{\bf 1}]
(\hat{\bf 1}-e^{i\omega'}{\hat\Gamma})^{-1}{\hat G}\}
\]
and with the expression of $F(\omega)$ in Eq. (\ref{generating_func00a})
\beq
=e^{i\omega'}
Tr\{[(\hat{\bf 1}-e^{i(\omega+\omega')}{\hat C}_2)^{-1}+(\hat{\bf 1}-e^{-i\omega}{\hat C}_1)^{-1}]
(\hat{\bf 1}-e^{i\omega'}{\hat\Gamma})^{-1}{\hat G}\}-F(\omega')
\ .
\label{generating_f3}
\eeq
To calculate the first term we can use the identity (\ref{identity7a}). 
With $A_1\times(A_2+B_2)=A_1\times A_2+A_1\times B_2$ we get the relation
\[
(\hat{\bf 1}-z{\hat C}_j)^{-1}=\cases{
[{\bf 1}-z\langle D^*\rangle_\tau({\bf 1}-E\Pi)]^{-1}\times{\bf 1} & $j=1$ \cr
{\bf 1}\times[{\bf 1}-z\langle D\rangle_\tau({\bf 1}-E\Pi)]^{-1} & $j=2$ \cr
}
\]
such that 
\[
[{\bf 1}-z\langle D^*\rangle_\tau({\bf 1}-E\Pi)]^{-1}
=[(z\langle D^*\rangle_\tau)^{-1}-{\bf 1}+E\Pi)]^{-1} (z\langle D^*\rangle_\tau)^{-1}
\ .
\]
This yields for the first term in Eq. (\ref{generating_f3})
\[
Tr\{({\bf 1}\times[{\bf 1}-e^{i(\omega+\omega')}\langle D\rangle_\tau({\bf 1}-E\Pi)]^{-1})(\hat{\bf 1}-e^{i\omega'}{\hat\Gamma})^{-1}{\hat G})\}
\]
\[
=Tr\{{\hat\Pi}({\bf 1}\times[{\bf 1}-e^{i(\omega+\omega')}\langle D\rangle_\tau({\bf 1}-E\Pi)]^{-1})(\hat{\bf 1}-e^{i\omega'}{\hat\Gamma})^{-1}
\langle{\hat D}\rangle_\tau{\hat E})\}
\]
\[
=Tr\{({\bf 1}\times[{\bf 1}-e^{i(\omega+\omega')}\langle D\rangle_\tau({\bf 1}-\Pi E)]^{-1}){\hat\Pi}(\hat{\bf 1}-e^{i\omega'}{\hat\Gamma})^{-1}
\langle{\hat D}\rangle_\tau{\hat E})\}
\]
\beq
=\sum_{j,j'}[{\bf 1}-e^{i(\omega+\omega')}\langle D\rangle_\tau({\bf 1}-\Pi E)]^{-1}_{j,j'}p_{j'}A_{j'}(\omega')
\label{trace2}
\eeq
with
\[
A_{j'}(\omega')=\sum_{j_1,j_2,j_3}[(\hat{\bf 1}-e^{i\omega'}{\hat\Gamma})^{-1}\langle D^*\times D\rangle_\tau]_{j_1 j',j_2j_3}p_{j_1}
\ .
\]
These components are linked to the generating function
$F(\omega')$ in Eq. (\ref{generating_func00a}) through $F(\omega')=\sum_j p_jA_j(\omega')$.
Now we write
\[
\sum_{j,j'}[{\bf 1}-e^{i(\omega+\omega')}\langle D\rangle_\tau({\bf 1}-\Pi E)]^{-1}_{j,j'}p_{j'}A_{j'}
=\sum_{j,j'}[(e^{i(\omega+\omega')}\langle D\rangle_\tau)^{-1}-{\bf 1}+\Pi E)]^{-1}_{j,j'}\frac{p_{j'}A_{j'}(\omega')}
{e^{i(\omega+\omega')}[\langle D\rangle_\tau]_{j'j'}}
\]
and with $K=(e^{i(\omega+\omega')}\langle D^*\rangle_\tau)^{-1}-{\bf 1}$ we can apply Eq. (\ref{identity7b})
\beq
=\sum_{j,j'}[K+\Pi E)]^{-1}_{j,j'}\frac{p_{j'}A_{j'}(\omega')}{e^{i(\omega+\omega')}[\langle D\rangle_\tau]_{j'j'}}
=e^{-i(\omega+\omega')}\frac{\sum_{j}K^{-1}_{j,j}p_{j}A_{j}(\omega')/\gamma_j^*}
{1+\sum_jK^{-1}_{j,j}p_j}
=\frac{\sum_jp_jA_j(\omega')\prod_{k\ne j}(1-e^{i(\omega+\omega')}\gamma_k^*)}{\sum_{j'}p_{j'}\prod_{k\ne j'}(1-e^{i(\omega+\omega')}\gamma_k^*)}
\label{iden_8a}
\eeq
The corresponding calculation yields for the second term in Eq. (\ref{generating_f3}):
\[
Tr\{([{\bf 1}-e^{-i\omega}\langle D^*\rangle_\tau({\bf 1}-E\Pi)]^{-1}
\times{\bf 1})(\hat{\bf 1}-e^{-i\omega'}{\hat\Gamma})^{-1}{\hat G})\}
=\sum_{j,j'}[{\bf 1}-e^{-i\omega}\langle D^*\rangle_\tau({\bf 1}-\Pi E)]^{-1}_{j,j'}p_{j'}B_{j'}(\omega')
\]
and again with Eq. (\ref{identity7b})
\beq
=\sum_{j,j'}(K'+\Pi E)^{-1}_{j,j'}\frac{p_{j'}B_{j'}(\omega')}{e^{-i\omega}[\langle D^*\rangle_\tau]_{j'j'}}
=e^{i\omega}\frac{\sum_{j}{K'}^{-1}_{j,j}p_{j}B_{j}(\omega')/\gamma_j}
{1+\sum_j{K'}^{-1}_{j,j}p_j}
=\frac{\sum_jp_jB_j(\omega')\prod_{k\ne j}(1-e^{-i\omega}\gamma_{k})}{\sum_{j'}p_{j'}\prod_{k\ne j'}(1-e^{-i\omega}\gamma_k)}
\label{iden_8b}
\eeq
with
\[
B_{j'}(\omega')=\sum_{j_1,j_2,j_3}[(\hat{\bf 1}-e^{-i\omega'}{\hat\Gamma})^{-1}\langle D^*\times D\rangle_\tau]_{j'j_1,j_2j_3}p_{j_1}
\ ,\ \ 
K'=(e^{-i\omega}\langle D\rangle_\tau)^{-1}-{\bf 1}
\]
and
$\sum_j p_jB_j(\omega')=F(\omega')$.
This allows us to write for the Fourier transform of Eq. (\ref{off-diag_phia})
\beq
\langle {\tilde\phi}^*(\omega){\tilde\phi}(\omega+\omega')\rangle_\tau
=e^{i\omega'}\sum_{j_1,j_2}p_{j_1}p_{j_2}(h_{j_2}+h'_{j_1})
\sum_{j_1',j_2'}\left[(\hat{\bf 1}-e^{i\omega'}{\hat\Gamma})^{-1}\langle{\hat D}\rangle_\tau\right]_{j_1j_2,j_1'j_2'}
-F(\omega')
\label{phiphi}
\eeq
with
\beq
h_j=\frac{\prod_{k\ne j}(1-e^{i(\omega+\omega')}\gamma_k^*)}{\sum_{j'}p_{j'}\prod_{k\ne j'}(1-e^{i(\omega+\omega')}\gamma_k^*)}
\ ,\ \ 
h_j'=\frac{\prod_{k\ne j}(1-e^{-i\omega}\gamma_{k})}{\sum_{j'}p_{j'}\prod_{k\ne j'}(1-e^{-i\omega}\gamma_k)}
\ ,\ \ \gamma_k=\langle e^{iE_k\tau}\rangle_\tau
\ .
\label{h_func}
\eeq
These are analytic functions in $e^{i(\omega+\omega')}$ and $e^{-i\omega}$, respectively, and their special form 
implies $\sum_jp_jh_j=\sum_jp_jh_j'=1$.
For $\omega'=0$ the normalization $\langle |{\tilde\phi}(\omega)|^2\rangle_\tau=1$ can be obtained
from Eq. (\ref{phiphi}) with the help of Eq. (\ref{sequence}). This can be shown by the following reasoning.
From App. \ref{app:norm} we have $F(\omega'=0)=1$. Then we can write with Eq. (\ref{phiphi})
\beq
\langle |{\tilde\phi}(\omega)|^2\rangle_\tau
=\sum_{j_1,j_2}p_{j_1}p_{j_2}(h_{j_2}+h'_{j_1})
\sum_{j_3,j_4}\left[(\hat{\bf 1}-{\hat\Gamma})^{-1}\langle{\hat D}\rangle_\tau
\right]_{j_1j_2,j_3j_4} -1
=\sum_{j_1,j_2}p_{j_1}p_{j_2}(h_{j_2}+h'_{j_1})T_{j_1j_2}-1
\ .
\label{norm2}
\eeq
Inserting now $T_{j_1j_2}$ from Eq. (\ref{sequence}) and use $\sum_jp_jh_j=\sum_jp_jh_j'=1$ 
we obtain 1 for this expression.

\section{Analytic properties of $F(\omega)$}
\label{app:gamma_an}

When we consider the trace term in Eq. (\ref{phi^2}) as
\beq
\langle|\phi_{n+1}|^2\rangle_\tau=Tr[(\hat{D}\hat{C})^n\hat{D}\hat{E}\hat{\Pi}]
\label{phi^2a}
\eeq
with the short-hand notation $\hat{D}=\langle\hat{D}\rangle_\tau$,
we get in the Zeno limit $\hat{D}\to\hat{\bf 1}$ a vanishing expression except for $n=0$,
since $\hat{E}\hat{\Pi}$ and $\hat{C}$ are projectors with
\[
(\hat{E}\hat{\Pi})^2=\hat{E}\hat{\Pi}
\ ,\ \ \
\hat{C}^2=\hat{C}
\ ,\ \ \
\hat{E}\hat{\Pi}\hat{C}=\hat{E}\hat{\Pi}({\bf 1}-E\Pi)\times({\bf 1}-E\Pi)=0
\ ,
\]
and since for $n=0$
\[
Tr(\hat{E}\hat{\Pi})=\sum_{j,j'=1}^Np_jp_{j'}=1
\ .
\] 
Returning to Eq. (\ref{phi^2a}), we can write with $\hat{C}^2=\hat{C}$ and $\hat{R}=\hat{\bf 1}-\hat{D}$
\beq
(\hat{D}\hat{C})^n=\hat{D}(\hat{C}\hat{D}\hat{C})^n
=\hat{D}[\hat{C}(\hat{\bf 1}-\hat{R})\hat{C}]^n
=\hat{D}(\hat{C}-\hat{C}\hat{R}\hat{C})^n
=\hat{D}\hat{C}(\hat{\bf 1}-\hat{R}\hat{C})^n
\ ,
\eeq
such that for $n\ge 1$
\[
\langle|\phi_{n+1}|^2\rangle_\tau=Tr[\hat{C}(\hat{\bf 1}-\hat{R}\hat{C})^n\hat{D}\hat{E}\hat{\Pi}\hat{D}]
=Tr[\hat{C}(\hat{\bf 1}-\hat{R}\hat{C})^n(\hat{\bf 1}-\hat{R})\hat{E}\hat{\Pi}(\hat{\bf 1}-\hat{R})]
\]
and with $\hat{E}\hat{\Pi}\hat{C}=\hat{C}\hat{E}\hat{\Pi}=0$
\beq
\langle|\phi_{n+1}|^2\rangle_\tau=Tr[\hat{C}(\hat{\bf 1}-\hat{R}\hat{C})^n\hat{R}\hat{E}\hat{\Pi}\hat{R}]
=Tr[\hat{R}\hat{C}(\hat{\bf 1}-\hat{R}\hat{C})^n\hat{R}\hat{E}\hat{\Pi}]
\ .
\eeq
Moreover, with $\hat{R}\hat{C}=\hat{\bf 1}-(\hat{\bf 1}-\hat{R}\hat{C})$ we get
\[
\langle|\phi_{n+1}|^2\rangle_\tau=Tr[(\hat{\bf 1}-\hat{R}\hat{C})^n\hat{R}\hat{E}\hat{\Pi}]
-Tr[(\hat{\bf 1}-\hat{R}\hat{C})^{n+1}\hat{R}\hat{E}\hat{\Pi}]
\ . 
\]
Now we introduce the projector $\hat{P}$ with $\hat{R}=\hat{P}\hat{R}$. Then we can write
\[
\hat{C}(\hat{\bf 1}-\hat{R}\hat{C})^n\hat{R}=\hat{C}(\hat{\bf 1}-\hat{R}\hat{C}\hat{P})^n\hat{R}
=\hat{C}(\hat{\bf 1}-\hat{P}+\hat{P}-\hat{R}\hat{C}\hat{P})^n\hat{R}
=\hat{C}[(\hat{\bf 1}-\hat{P})^n+(\hat{P}-\hat{R}\hat{C}\hat{P})^n]\hat{R}
\]
and since $(\hat{\bf 1}-\hat{P})^n\hat{R}=0$
\beq
=\hat{C}(\hat{P}-\hat{R}\hat{C}\hat{P})^n\hat{R}
\ .
\eeq
The eigenvalues of $\hat{R}\hat{C}\hat{P}$ might be complex. Therefore, it is better to calculate the 
eigenvalues of the Hermitean matrix
\[
(\hat{R}\hat{C}\hat{P})^\dagger\hat{R}\hat{C}\hat{P}
=\hat{P}\hat{C}\hat{R}^\dagger\hat{R}\hat{C}\hat{P}
\ ,
\]
whose determinant reads
\beq
\det(\hat{P}\hat{C}\hat{R}^\dagger\hat{R}\hat{C}\hat{P})
=\det(\hat{P}\hat{C}\hat{P})^2\prod_{j,j'=1;j'\ne j}^N|1-\langle D_{jj'}\rangle_\tau|^2
\ .
\eeq
A necessary condition for a quick decay of $(\hat{P}-\hat{R}\hat{C}\hat{P})^n$ with $n$ is that 
the product of $|1-\langle D_{jj'}\rangle_\tau|^2$ is not small, while the sufficient condition 
requires that $\det(\hat{P}\hat{C}\hat{P})$ also is not small.
To see the latter, we analyze the elements of the projected matrix 
\beq
(\hat{P}\hat{C}\hat{P})_{jj',kk'}=(1-\delta_{jj'})(1-\delta_{kk'})
(\delta_{jk}-p_k)(\delta_{j'k'}-p_{k'})
\ .
\eeq
We only consider the projected matrix, which has the following matrix elements with $j\ne j'$ 
and $k\ne k'$:
\[
k=j, k'=j':\ \
(\hat{P}\hat{C}\hat{P})_{jj',jj'}=(1-p_j)(1-p_{j'})
\ ,\ \ 
k\ne j, k'\ne j': \ \ (\hat{P}\hat{C}\hat{P})_{jj',kk'}=p_kp_{k'}
\]
and 
\[
k=j, k'\ne j' :\ \ 
(\hat{P}\hat{C}\hat{P})_{jj',jk'}=-(1-p_j)p_{k'}
\ ,\ \ 
k\ne j, k'= j': \ \ (\hat{P}\hat{C}\hat{P})_{jj',kj'}=-p_k(1-p_{j'})
\ .
\]
For the special case of $N=2$ this gives a $2\times 2$ matrix:
\[
\pmatrix{
0 & -p_2(1-p_2) \cr
-p_1(1-p_1) & 0 \cr
}
\ ,
\]
whose determinant $-p_1(1-p_1)p_2(1-p_2)$ vanishes only for $p_1=0,1$ and/or $p_2=0,1$.

\section{Symmetric two-level system}
\label{app:2ls}

The matrix structure of the symmetric 2LS reads
\[
([A\times B]_{ij,kl})=(A_{ik}B_{jl})
=\pmatrix{
A_{11}B_{11} & A_{11}B_{12} & A_{12}B_{11} & A_{12}B_{12}\cr
A_{11}B_{21} & A_{11}B_{22} & A_{12}B_{21} & A_{12}B_{22}\cr
A_{21}B_{11} & A_{21}B_{12} & A_{22}B_{11} & A_{22}B_{12}\cr
A_{21}B_{21} & A_{21}B_{22} & A_{22}B_{21} & A_{22}B_{22}\cr
}
\]
and
\[
(\Gamma_{ij,kl})
=\pmatrix{
\Gamma_{11,11} & \Gamma_{11,12} & \Gamma_{11,21} & \Gamma_{11,22} \cr
\Gamma_{12,11} & \Gamma_{12,12} & \Gamma_{12,21} & \Gamma_{12,22} \cr
\Gamma_{21,11} & \Gamma_{21,12} & \Gamma_{21,21} & \Gamma_{21,22} \cr
\Gamma_{22,11} & \Gamma_{22,12} & \Gamma_{22,21} & \Gamma_{22,22} \cr
}
\]
Then we have
\[
\langle D^*\times D\rangle_\tau=\pmatrix{
1 & 0 & 0 & 0 \cr
0 & \langle e^{2iJ\tau}\rangle_\tau & 0 & 0 \cr
0 & 0 &  \langle e^{-2iJ\tau}\rangle_\tau & 0 \cr
0 & 0 & 0 & 1 \cr
}
=\pmatrix{
1 & 0 & 0 & 0 \cr
0 & y & 0 & 0 \cr
0 & 0 & y^* & 0 \cr
0 & 0 & 0 & 1 \cr
}
\]
with $y=\langle e^{2iJ\tau}\rangle_\tau$. Moreover, 
\[
({\bf 1}-E\Pi)\times ({\bf 1}-E\Pi)
=\frac{1}{4}\pmatrix{
1 & -1 & -1 & 1\cr
-1 & 1 & 1 & -1\cr
-1 & 1 & 1 & -1\cr
1 & -1 & -1 & 1\cr
}
\]
such that
\[
{\hat\Gamma}=\langle D^*\times D\rangle_\tau ({\bf 1}-E\Pi)\times ({\bf 1}-E\Pi)
=\frac{1}{4}\pmatrix{
-1 & 1 & 1 & -1\cr
y & -y & -y & y\cr
y^* & -y^* & -y^* & y^*\cr
-1 & 1 & 1 & -1\cr
}
\]
with three vanishing eigenvalues and one eigenvalue $(y+y^*+2)/4$.
With the help of Maxima 
we obtain with $c=z/4$
\[
\det(\hat{\bf 1}-z{\hat\Gamma})=1-c(2+y+y^*)=1-\frac{z}{2}[1+\langle\cos(2J \tau)\rangle_\tau]
=1-z\langle\cos^2(J \tau)\rangle_\tau
\]
\[
(\hat{\bf 1}-z{\hat\Gamma})^{-1}
=\frac{-1}{1-z\langle\cos^2(J \tau)\rangle_\tau}
\pmatrix{
c(y+y^*+1)-1 & c & c & -c \cr
cy & c(y^*+2)-1 & -cy & cy \cr
cy^* & -cy^* & c(y+2)-1 & cy^* \cr
-c & c & c & c(y+y^*+1)-1 \cr
}
\]
\[
(\hat{\bf 1}-z{\hat\Gamma})^{-1}\langle D^*\times D\rangle_\tau
\]
\[
=\frac{-1}{1-z\langle\cos^2(J \tau)\rangle_\tau}
\pmatrix{
c(y+y^*+1)-1 & cy & cy^* & -c \cr
cy & c(yy^*+2y)-y & -cyy^* & cy \cr
cy^* & -cyy^* & c(yy^*+2y^*)-y^* & cy^* \cr
-c & cy & cy^* & c(y+y^*+1)-1 \cr
}
\]
Then the generating function $F(\omega)$ in Eq. (\ref{generating_func00a}) reads
\[
F(\omega)
=\frac{4e^{2i\omega}\langle\cos2J\tau\rangle_\tau-2e^{i\omega}(\langle\cos2J\tau\rangle_\tau+1)}{2e^{i\omega}(\langle\cos2J\tau\rangle_\tau+1)-4}
\ ,
\]
which gives for $\omega=0$
\[
F(0)=1
\ ,\ \
-iF'(0)=2
\ ,\ \ 
-F''(0)=2\frac{3-\langle\cos2J\tau\rangle_\tau}{1-\langle\cos2J\tau\rangle_\tau}
\ .
\]
For the $\phi$--correlator we get the winding number
\beq
w_\phi=
\frac{1}{2\pi i}\int_0^{2\pi}\partial_{\omega'}\log(\langle {\tilde\phi}^*(\omega){\tilde\phi}(\omega+\omega')\rangle_\tau)\Big|_{\omega'=0}d\omega
=\frac{1}{2\pi i}\int\frac{a_2z^2+a_1z+a_0}{4(y+y^*-2)(z-C)(Cz-1)}\frac{1}{z}dz
\label{wind_n2}
\eeq
with $C=
\langle\cos J\tau\rangle_\tau$ and $y=\langle e^{2iJ\tau}\rangle_\tau$, with poles
\[
z_0=0
\ ,\ \
z_1=C
\ ,\ \ 
z_2=1/C
\]
and with the coefficients
\[
a_0=a_2=(6x^*+2x)y^*+(2x^*+ 6x)y-8x^*-8x
\]
\[
a_1=4[(-{x^*}^2-xx^*-2)y^*+(-xx^*-x^2-2)y+{x^*}^2+2xx^*+x^2+4]
\ ,
\]
where $x=\langle e^{-iJ\tau}\rangle_\tau$.
After performing the Cauchy integration in Eq. (\ref{wind_n2}) for the two poles $z_{0,1}$ we get
\[
w_\phi=2
\ .
\]

\bibliography{rand_time_2.bib}

\begin{thebibliography}{27}%
\makeatletter
\providecommand \@ifxundefined [1]{%
 \@ifx{#1\undefined}
}%
\providecommand \@ifnum [1]{%
 \ifnum #1\expandafter \@firstoftwo
 \else \expandafter \@secondoftwo
 \fi
}%
\providecommand \@ifx [1]{%
 \ifx #1\expandafter \@firstoftwo
 \else \expandafter \@secondoftwo
 \fi
}%
\providecommand \natexlab [1]{#1}%
\providecommand \enquote  [1]{``#1''}%
\providecommand \bibnamefont  [1]{#1}%
\providecommand \bibfnamefont [1]{#1}%
\providecommand \citenamefont [1]{#1}%
\providecommand \href@noop [0]{\@secondoftwo}%
\providecommand \href [0]{\begingroup \@sanitize@url \@href}%
\providecommand \@href[1]{\@@startlink{#1}\@@href}%
\providecommand \@@href[1]{\endgroup#1\@@endlink}%
\providecommand \@sanitize@url [0]{\catcode `\\12\catcode `\$12\catcode
  `\&12\catcode `\#12\catcode `\^12\catcode `\_12\catcode `\%12\relax}%
\providecommand \@@startlink[1]{}%
\providecommand \@@endlink[0]{}%
\providecommand \url  [0]{\begingroup\@sanitize@url \@url }%
\providecommand \@url [1]{\endgroup\@href {#1}{\urlprefix }}%
\providecommand \urlprefix  [0]{URL }%
\providecommand \Eprint [0]{\href }%
\providecommand \doibase [0]{https://doi.org/}%
\providecommand \selectlanguage [0]{\@gobble}%
\providecommand \bibinfo  [0]{\@secondoftwo}%
\providecommand \bibfield  [0]{\@secondoftwo}%
\providecommand \translation [1]{[#1]}%
\providecommand \BibitemOpen [0]{}%
\providecommand \bibitemStop [0]{}%
\providecommand \bibitemNoStop [0]{.\EOS\space}%
\providecommand \EOS [0]{\spacefactor3000\relax}%
\providecommand \BibitemShut  [1]{\csname bibitem#1\endcsname}%
\let\auto@bib@innerbib\@empty
\bibitem [{\citenamefont {\ifmmode \check{S}\else
  \v{S}\fi{}tefa\ifmmode~\check{n}\else \v{n}\fi{}\'ak}\ \emph
  {et~al.}(2008)\citenamefont {\ifmmode \check{S}\else
  \v{S}\fi{}tefa\ifmmode~\check{n}\else \v{n}\fi{}\'ak}, \citenamefont {Jex},\
  and\ \citenamefont {Kiss}}]{Stefanak2008}%
  \BibitemOpen
  \bibfield  {author} {\bibinfo {author} {\bibfnamefont {M.}~\bibnamefont
  {\ifmmode \check{S}\else \v{S}\fi{}tefa\ifmmode~\check{n}\else
  \v{n}\fi{}\'ak}}, \bibinfo {author} {\bibfnamefont {I.}~\bibnamefont {Jex}},\
  and\ \bibinfo {author} {\bibfnamefont {T.}~\bibnamefont {Kiss}},\ }\bibfield
  {title} {\bibinfo {title} {Recurrence and p\'olya number of quantum walks},\
  }\href {https://doi.org/10.1103/PhysRevLett.100.020501} {\bibfield  {journal}
  {\bibinfo  {journal} {Phys. Rev. Lett.}\ }\textbf {\bibinfo {volume} {100}},\
  \bibinfo {pages} {020501} (\bibinfo {year} {2008})}\BibitemShut {NoStop}%
\bibitem [{\citenamefont {Bach}\ \emph {et~al.}(2004)\citenamefont {Bach},
  \citenamefont {Coppersmith}, \citenamefont {Goldschen}, \citenamefont
  {Joynt},\ and\ \citenamefont {Watrous}}]{bach04}%
  \BibitemOpen
  \bibfield  {author} {\bibinfo {author} {\bibfnamefont {E.}~\bibnamefont
  {Bach}}, \bibinfo {author} {\bibfnamefont {S.}~\bibnamefont {Coppersmith}},
  \bibinfo {author} {\bibfnamefont {M.~P.}\ \bibnamefont {Goldschen}}, \bibinfo
  {author} {\bibfnamefont {R.}~\bibnamefont {Joynt}},\ and\ \bibinfo {author}
  {\bibfnamefont {J.}~\bibnamefont {Watrous}},\ }\bibfield  {title} {\bibinfo
  {title} {One-dimensional quantum walks with absorbing boundaries},\ }\href
  {https://doi.org/https://doi.org/10.1016/j.jcss.2004.03.005} {\bibfield
  {journal} {\bibinfo  {journal} {Journal of Computer and System Sciences}\
  }\textbf {\bibinfo {volume} {69}},\ \bibinfo {pages} {562 } (\bibinfo {year}
  {2004})}\BibitemShut {NoStop}%
\bibitem [{\citenamefont {Friedman}\ \emph {et~al.}(2017)\citenamefont
  {Friedman}, \citenamefont {Kessler},\ and\ \citenamefont
  {Barkai}}]{Friedman2017}%
  \BibitemOpen
  \bibfield  {author} {\bibinfo {author} {\bibfnamefont {H.}~\bibnamefont
  {Friedman}}, \bibinfo {author} {\bibfnamefont {D.~A.}\ \bibnamefont
  {Kessler}},\ and\ \bibinfo {author} {\bibfnamefont {E.}~\bibnamefont
  {Barkai}},\ }\bibfield  {title} {\bibinfo {title} {Quantum walks: The first
  detected passage time problem},\ }\href
  {https://doi.org/10.1103/PhysRevE.95.032141} {\bibfield  {journal} {\bibinfo
  {journal} {Phys. Rev. E}\ }\textbf {\bibinfo {volume} {95}},\ \bibinfo
  {pages} {032141} (\bibinfo {year} {2017})}\BibitemShut {NoStop}%
\bibitem [{\citenamefont {Dhar}\ \emph
  {et~al.}(2015{\natexlab{a}})\citenamefont {Dhar}, \citenamefont {Dasgupta},
  \citenamefont {Dhar},\ and\ \citenamefont {Sen}}]{dhar15}%
  \BibitemOpen
  \bibfield  {author} {\bibinfo {author} {\bibfnamefont {S.}~\bibnamefont
  {Dhar}}, \bibinfo {author} {\bibfnamefont {S.}~\bibnamefont {Dasgupta}},
  \bibinfo {author} {\bibfnamefont {A.}~\bibnamefont {Dhar}},\ and\ \bibinfo
  {author} {\bibfnamefont {D.}~\bibnamefont {Sen}},\ }\bibfield  {title}
  {\bibinfo {title} {Detection of a quantum particle on a lattice under
  repeated projective measurements},\ }\href
  {https://doi.org/10.1103/PhysRevA.91.062115} {\bibfield  {journal} {\bibinfo
  {journal} {Phys. Rev. A}\ }\textbf {\bibinfo {volume} {91}},\ \bibinfo
  {pages} {062115} (\bibinfo {year} {2015}{\natexlab{a}})}\BibitemShut
  {NoStop}%
\bibitem [{\citenamefont {Dhar}\ \emph
  {et~al.}(2015{\natexlab{b}})\citenamefont {Dhar}, \citenamefont {Dasgupta},\
  and\ \citenamefont {Dhar}}]{Dhar_2015}%
  \BibitemOpen
  \bibfield  {author} {\bibinfo {author} {\bibfnamefont {S.}~\bibnamefont
  {Dhar}}, \bibinfo {author} {\bibfnamefont {S.}~\bibnamefont {Dasgupta}},\
  and\ \bibinfo {author} {\bibfnamefont {A.}~\bibnamefont {Dhar}},\ }\bibfield
  {title} {\bibinfo {title} {Quantum time of arrival distribution in a simple
  lattice model},\ }\href {https://doi.org/10.1088/1751-8113/48/11/115304}
  {\bibfield  {journal} {\bibinfo  {journal} {Journal of Physics A:
  Mathematical and Theoretical}\ }\textbf {\bibinfo {volume} {48}},\ \bibinfo
  {pages} {115304} (\bibinfo {year} {2015}{\natexlab{b}})}\BibitemShut
  {NoStop}%
\bibitem [{\citenamefont {Krovi}\ and\ \citenamefont
  {Brun}(2006{\natexlab{a}})}]{krovi06}%
  \BibitemOpen
  \bibfield  {author} {\bibinfo {author} {\bibfnamefont {H.}~\bibnamefont
  {Krovi}}\ and\ \bibinfo {author} {\bibfnamefont {T.~A.}\ \bibnamefont
  {Brun}},\ }\bibfield  {title} {\bibinfo {title} {Quantum walks with infinite
  hitting times},\ }\href {https://doi.org/10.1103/PhysRevA.74.042334}
  {\bibfield  {journal} {\bibinfo  {journal} {Phys. Rev. A}\ }\textbf {\bibinfo
  {volume} {74}},\ \bibinfo {pages} {042334} (\bibinfo {year}
  {2006}{\natexlab{a}})}\BibitemShut {NoStop}%
\bibitem [{\citenamefont {Krovi}\ and\ \citenamefont
  {Brun}(2006{\natexlab{b}})}]{krovi06a}%
  \BibitemOpen
  \bibfield  {author} {\bibinfo {author} {\bibfnamefont {H.}~\bibnamefont
  {Krovi}}\ and\ \bibinfo {author} {\bibfnamefont {T.~A.}\ \bibnamefont
  {Brun}},\ }\bibfield  {title} {\bibinfo {title} {Hitting time for quantum
  walks on the hypercube},\ }\href {https://doi.org/10.1103/PhysRevA.73.032341}
  {\bibfield  {journal} {\bibinfo  {journal} {Phys. Rev. A}\ }\textbf {\bibinfo
  {volume} {73}},\ \bibinfo {pages} {032341} (\bibinfo {year}
  {2006}{\natexlab{b}})}\BibitemShut {NoStop}%
\bibitem [{\citenamefont {Krovi}\ and\ \citenamefont {Brun}(2007)}]{krovi07}%
  \BibitemOpen
  \bibfield  {author} {\bibinfo {author} {\bibfnamefont {H.}~\bibnamefont
  {Krovi}}\ and\ \bibinfo {author} {\bibfnamefont {T.~A.}\ \bibnamefont
  {Brun}},\ }\bibfield  {title} {\bibinfo {title} {Quantum walks on quotient
  graphs},\ }\href {https://doi.org/10.1103/PhysRevA.75.062332} {\bibfield
  {journal} {\bibinfo  {journal} {Phys. Rev. A}\ }\textbf {\bibinfo {volume}
  {75}},\ \bibinfo {pages} {062332} (\bibinfo {year} {2007})}\BibitemShut
  {NoStop}%
\bibitem [{\citenamefont {Gr{\"u}nbaum}\ \emph {et~al.}(2013)\citenamefont
  {Gr{\"u}nbaum}, \citenamefont {Vel{\'a}zquez}, \citenamefont {Werner},\ and\
  \citenamefont {Werner}}]{Gruenbaum2013}%
  \BibitemOpen
  \bibfield  {author} {\bibinfo {author} {\bibfnamefont {F.~A.}\ \bibnamefont
  {Gr{\"u}nbaum}}, \bibinfo {author} {\bibfnamefont {L.}~\bibnamefont
  {Vel{\'a}zquez}}, \bibinfo {author} {\bibfnamefont {A.~H.}\ \bibnamefont
  {Werner}},\ and\ \bibinfo {author} {\bibfnamefont {R.~F.}\ \bibnamefont
  {Werner}},\ }\bibfield  {title} {\bibinfo {title} {Recurrence for discrete
  time unitary evolutions},\ }\href {https://doi.org/10.1007/s00220-012-1645-2}
  {\bibfield  {journal} {\bibinfo  {journal} {Communications in Mathematical
  Physics}\ }\textbf {\bibinfo {volume} {320}},\ \bibinfo {pages} {543}
  (\bibinfo {year} {2013})}\BibitemShut {NoStop}%
\bibitem [{\citenamefont {Krapivsky}\ \emph {et~al.}(2014)\citenamefont
  {Krapivsky}, \citenamefont {Luck},\ and\ \citenamefont {Mallick}}]{luck14}%
  \BibitemOpen
  \bibfield  {author} {\bibinfo {author} {\bibfnamefont {P.}~\bibnamefont
  {Krapivsky}}, \bibinfo {author} {\bibfnamefont {J.}~\bibnamefont {Luck}},\
  and\ \bibinfo {author} {\bibfnamefont {K.}~\bibnamefont {Mallick}},\
  }\bibfield  {title} {\bibinfo {title} {Survival of classical and quantum
  particles in the presence of traps},\ }\href
  {https://doi.org/10.1007/s10955-014-0936-8} {\bibfield  {journal} {\bibinfo
  {journal} {J Stat Phys}\ }\textbf {\bibinfo {volume} {154}},\ \bibinfo
  {pages} {1430} (\bibinfo {year} {2014})}\BibitemShut {NoStop}%
\bibitem [{\citenamefont {Bourgain}\ \emph {et~al.}(2014)\citenamefont
  {Bourgain}, \citenamefont {Gr{\"u}nbaum}, \citenamefont {Vel{\'a}zquez},\
  and\ \citenamefont {Wilkening}}]{Grunbaum2014}%
  \BibitemOpen
  \bibfield  {author} {\bibinfo {author} {\bibfnamefont {J.}~\bibnamefont
  {Bourgain}}, \bibinfo {author} {\bibfnamefont {F.~A.}\ \bibnamefont
  {Gr{\"u}nbaum}}, \bibinfo {author} {\bibfnamefont {L.}~\bibnamefont
  {Vel{\'a}zquez}},\ and\ \bibinfo {author} {\bibfnamefont {J.}~\bibnamefont
  {Wilkening}},\ }\bibfield  {title} {\bibinfo {title} {Quantum recurrence of a
  subspace and operator-valued schur functions},\ }\href
  {https://doi.org/10.1007/s00220-014-1929-9} {\bibfield  {journal} {\bibinfo
  {journal} {Communications in Mathematical Physics}\ }\textbf {\bibinfo
  {volume} {329}},\ \bibinfo {pages} {1031} (\bibinfo {year}
  {2014})}\BibitemShut {NoStop}%
\bibitem [{\citenamefont {Sinkovicz}\ \emph {et~al.}(2016)\citenamefont
  {Sinkovicz}, \citenamefont {Kiss},\ and\ \citenamefont
  {Asb\'oth}}]{sinkovicz16}%
  \BibitemOpen
  \bibfield  {author} {\bibinfo {author} {\bibfnamefont {P.}~\bibnamefont
  {Sinkovicz}}, \bibinfo {author} {\bibfnamefont {T.}~\bibnamefont {Kiss}},\
  and\ \bibinfo {author} {\bibfnamefont {J.~K.}\ \bibnamefont {Asb\'oth}},\
  }\bibfield  {title} {\bibinfo {title} {Generalized kac lemma for recurrence
  time in iterated open quantum systems},\ }\href
  {https://doi.org/10.1103/PhysRevA.93.050101} {\bibfield  {journal} {\bibinfo
  {journal} {Phys. Rev. A}\ }\textbf {\bibinfo {volume} {93}},\ \bibinfo
  {pages} {050101} (\bibinfo {year} {2016})}\BibitemShut {NoStop}%
\bibitem [{\citenamefont {Friedman}\ \emph {et~al.}(2016)\citenamefont
  {Friedman}, \citenamefont {Kessler},\ and\ \citenamefont
  {Barkai}}]{Friedman_2016}%
  \BibitemOpen
  \bibfield  {author} {\bibinfo {author} {\bibfnamefont {H.}~\bibnamefont
  {Friedman}}, \bibinfo {author} {\bibfnamefont {D.~A.}\ \bibnamefont
  {Kessler}},\ and\ \bibinfo {author} {\bibfnamefont {E.}~\bibnamefont
  {Barkai}},\ }\bibfield  {title} {\bibinfo {title} {Quantum renewal equation
  for the first detection time of a quantum walk},\ }\href
  {https://doi.org/10.1088/1751-8121/aa5191} {\bibfield  {journal} {\bibinfo
  {journal} {Journal of Physics A: Mathematical and Theoretical}\ }\textbf
  {\bibinfo {volume} {50}},\ \bibinfo {pages} {04LT01} (\bibinfo {year}
  {2016})}\BibitemShut {NoStop}%
\bibitem [{\citenamefont {Thiel}\ \emph {et~al.}(2018)\citenamefont {Thiel},
  \citenamefont {Barkai},\ and\ \citenamefont {Kessler}}]{thiel18}%
  \BibitemOpen
  \bibfield  {author} {\bibinfo {author} {\bibfnamefont {F.}~\bibnamefont
  {Thiel}}, \bibinfo {author} {\bibfnamefont {E.}~\bibnamefont {Barkai}},\ and\
  \bibinfo {author} {\bibfnamefont {D.~A.}\ \bibnamefont {Kessler}},\
  }\bibfield  {title} {\bibinfo {title} {First detected arrival of a quantum
  walker on an infinite line},\ }\href
  {https://doi.org/10.1103/PhysRevLett.120.040502} {\bibfield  {journal}
  {\bibinfo  {journal} {Phys. Rev. Lett.}\ }\textbf {\bibinfo {volume} {120}},\
  \bibinfo {pages} {040502} (\bibinfo {year} {2018})}\BibitemShut {NoStop}%
\bibitem [{\citenamefont {Nitsche}\ \emph {et~al.}(2018)\citenamefont
  {Nitsche}, \citenamefont {Barkhofen}, \citenamefont {Kruse}, \citenamefont
  {Sansoni}, \citenamefont {{\v S}tefa{\v n}{\'a}k}, \citenamefont
  {G{\'a}bris}, \citenamefont {Poto{\v c}ek}, \citenamefont {Kiss},
  \citenamefont {Jex},\ and\ \citenamefont {Silberhorn}}]{nitsche18}%
  \BibitemOpen
  \bibfield  {author} {\bibinfo {author} {\bibfnamefont {T.}~\bibnamefont
  {Nitsche}}, \bibinfo {author} {\bibfnamefont {S.}~\bibnamefont {Barkhofen}},
  \bibinfo {author} {\bibfnamefont {R.}~\bibnamefont {Kruse}}, \bibinfo
  {author} {\bibfnamefont {L.}~\bibnamefont {Sansoni}}, \bibinfo {author}
  {\bibfnamefont {M.}~\bibnamefont {{\v S}tefa{\v n}{\'a}k}}, \bibinfo {author}
  {\bibfnamefont {A.}~\bibnamefont {G{\'a}bris}}, \bibinfo {author}
  {\bibfnamefont {V.}~\bibnamefont {Poto{\v c}ek}}, \bibinfo {author}
  {\bibfnamefont {T.}~\bibnamefont {Kiss}}, \bibinfo {author} {\bibfnamefont
  {I.}~\bibnamefont {Jex}},\ and\ \bibinfo {author} {\bibfnamefont
  {C.}~\bibnamefont {Silberhorn}},\ }\bibfield  {title} {\bibinfo {title}
  {Probing measurement-induced effects in quantum walks via recurrence},\
  }\bibfield  {journal} {\bibinfo  {journal} {Science Advances}\ }\textbf
  {\bibinfo {volume} {4}},\ \href {https://doi.org/10.1126/sciadv.aar6444}
  {10.1126/sciadv.aar6444} (\bibinfo {year} {2018}),\ \Eprint
  {https://arxiv.org/abs/https://advances.sciencemag.org/content/4/6/eaar6444.full.pdf}
  {https://advances.sciencemag.org/content/4/6/eaar6444.full.pdf} \BibitemShut
  {NoStop}%
\bibitem [{\citenamefont {{Krapivsky}}\ \emph {et~al.}(2018)\citenamefont
  {{Krapivsky}}, \citenamefont {{Luck}},\ and\ \citenamefont
  {{Mallick}}}]{2018JSMTE..02.3104K}%
  \BibitemOpen
  \bibfield  {author} {\bibinfo {author} {\bibfnamefont {P.~L.}\ \bibnamefont
  {{Krapivsky}}}, \bibinfo {author} {\bibfnamefont {J.~M.}\ \bibnamefont
  {{Luck}}},\ and\ \bibinfo {author} {\bibfnamefont {K.}~\bibnamefont
  {{Mallick}}},\ }\bibfield  {title} {\bibinfo {title} {{Quantum return
  probability of a system of N non-interacting lattice fermions}},\ }\href
  {https://doi.org/10.1088/1742-5468/aaa79a} {\bibfield  {journal} {\bibinfo
  {journal} {Journal of Statistical Mechanics: Theory and Experiment}\ }\textbf
  {\bibinfo {volume} {2}},\ \bibinfo {pages} {023104} (\bibinfo {year}
  {2018})},\ \Eprint {https://arxiv.org/abs/1710.08178} {arXiv:1710.08178
  [cond-mat.mes-hall]} \BibitemShut {NoStop}%
\bibitem [{\citenamefont {Lahiri}\ and\ \citenamefont {Dhar}(2019)}]{lahiri19}%
  \BibitemOpen
  \bibfield  {author} {\bibinfo {author} {\bibfnamefont {S.}~\bibnamefont
  {Lahiri}}\ and\ \bibinfo {author} {\bibfnamefont {A.}~\bibnamefont {Dhar}},\
  }\bibfield  {title} {\bibinfo {title} {Return to the origin problem for a
  particle on a one-dimensional lattice with quasi-zeno dynamics},\ }\href
  {https://doi.org/10.1103/PhysRevA.99.012101} {\bibfield  {journal} {\bibinfo
  {journal} {Phys. Rev. A}\ }\textbf {\bibinfo {volume} {99}},\ \bibinfo
  {pages} {012101} (\bibinfo {year} {2019})}\BibitemShut {NoStop}%
\bibitem [{\citenamefont {Yin}\ \emph {et~al.}(2019)\citenamefont {Yin},
  \citenamefont {Ziegler}, \citenamefont {Thiel},\ and\ \citenamefont
  {Barkai}}]{Yin2019}%
  \BibitemOpen
  \bibfield  {author} {\bibinfo {author} {\bibfnamefont {R.}~\bibnamefont
  {Yin}}, \bibinfo {author} {\bibfnamefont {K.}~\bibnamefont {Ziegler}},
  \bibinfo {author} {\bibfnamefont {F.}~\bibnamefont {Thiel}},\ and\ \bibinfo
  {author} {\bibfnamefont {E.}~\bibnamefont {Barkai}},\ }\bibfield  {title}
  {\bibinfo {title} {Large fluctuations of the first detected quantum return
  time},\ }\href {https://doi.org/10.1103/PhysRevResearch.1.033086} {\bibfield
  {journal} {\bibinfo  {journal} {Phys. Rev. Research}\ }\textbf {\bibinfo
  {volume} {1}},\ \bibinfo {pages} {033086} (\bibinfo {year}
  {2019})}\BibitemShut {NoStop}%
\bibitem [{\citenamefont {{Dubey}}\ \emph {et~al.}(2020)\citenamefont
  {{Dubey}}, \citenamefont {{Bernardin}},\ and\ \citenamefont
  {{Dhar}}}]{2020arXiv201201196D}%
  \BibitemOpen
  \bibfield  {author} {\bibinfo {author} {\bibfnamefont {V.}~\bibnamefont
  {{Dubey}}}, \bibinfo {author} {\bibfnamefont {C.}~\bibnamefont
  {{Bernardin}}},\ and\ \bibinfo {author} {\bibfnamefont {A.}~\bibnamefont
  {{Dhar}}},\ }\bibfield  {title} {\bibinfo {title} {{Quantum Dynamics under
  continuous projective measurements: non-Hermitian description and the
  continuous space space limit}},\ }\href@noop {} {\bibfield  {journal}
  {\bibinfo  {journal} {arXiv e-prints}\ ,\ \bibinfo {eid} {arXiv:2012.01196}}
  (\bibinfo {year} {2020})},\ \Eprint {https://arxiv.org/abs/2012.01196}
  {arXiv:2012.01196 [quant-ph]} \BibitemShut {NoStop}%
\bibitem [{\citenamefont {Liu}\ \emph {et~al.}(2020)\citenamefont {Liu},
  \citenamefont {Yin}, \citenamefont {Ziegler},\ and\ \citenamefont
  {Barkai}}]{quancheng20}%
  \BibitemOpen
  \bibfield  {author} {\bibinfo {author} {\bibfnamefont {Q.}~\bibnamefont
  {Liu}}, \bibinfo {author} {\bibfnamefont {R.}~\bibnamefont {Yin}}, \bibinfo
  {author} {\bibfnamefont {K.}~\bibnamefont {Ziegler}},\ and\ \bibinfo {author}
  {\bibfnamefont {E.}~\bibnamefont {Barkai}},\ }\bibfield  {title} {\bibinfo
  {title} {Quantum walks: The mean first detected transition time},\ }\href
  {https://doi.org/10.1103/PhysRevResearch.2.033113} {\bibfield  {journal}
  {\bibinfo  {journal} {Phys. Rev. Research}\ }\textbf {\bibinfo {volume}
  {2}},\ \bibinfo {pages} {033113} (\bibinfo {year} {2020})}\BibitemShut
  {NoStop}%
\bibitem [{\citenamefont {Varbanov}\ \emph {et~al.}(2008)\citenamefont
  {Varbanov}, \citenamefont {Krovi},\ and\ \citenamefont {Brun}}]{varbanov08}%
  \BibitemOpen
  \bibfield  {author} {\bibinfo {author} {\bibfnamefont {M.}~\bibnamefont
  {Varbanov}}, \bibinfo {author} {\bibfnamefont {H.}~\bibnamefont {Krovi}},\
  and\ \bibinfo {author} {\bibfnamefont {T.~A.}\ \bibnamefont {Brun}},\
  }\bibfield  {title} {\bibinfo {title} {Hitting time for the continuous
  quantum walk},\ }\href {https://doi.org/10.1103/PhysRevA.78.022324}
  {\bibfield  {journal} {\bibinfo  {journal} {Phys. Rev. A}\ }\textbf {\bibinfo
  {volume} {78}},\ \bibinfo {pages} {022324} (\bibinfo {year}
  {2008})}\BibitemShut {NoStop}%
\bibitem [{\citenamefont {{Riera-Campeny}}\ \emph {et~al.}(2020)\citenamefont
  {{Riera-Campeny}}, \citenamefont {{Oll{\'e}}},\ and\ \citenamefont
  {{Mas{\'o}-Puigdellosas}}}]{2020arXiv201104403R}%
  \BibitemOpen
  \bibfield  {author} {\bibinfo {author} {\bibfnamefont {A.}~\bibnamefont
  {{Riera-Campeny}}}, \bibinfo {author} {\bibfnamefont {J.}~\bibnamefont
  {{Oll{\'e}}}},\ and\ \bibinfo {author} {\bibfnamefont {A.}~\bibnamefont
  {{Mas{\'o}-Puigdellosas}}},\ }\bibfield  {title} {\bibinfo {title}
  {{Measurement-induced resetting in open quantum systems}},\ }\href@noop {}
  {\bibfield  {journal} {\bibinfo  {journal} {arXiv e-prints}\ ,\ \bibinfo
  {eid} {arXiv:2011.04403}} (\bibinfo {year} {2020})},\ \Eprint
  {https://arxiv.org/abs/2011.04403} {arXiv:2011.04403 [quant-ph]} \BibitemShut
  {NoStop}%
\bibitem [{\citenamefont {Redner}(2001)}]{Redner2001}%
  \BibitemOpen
  \bibfield  {author} {\bibinfo {author} {\bibfnamefont {S.}~\bibnamefont
  {Redner}},\ }\href {https://doi.org/10.1017/CBO9780511606014} {\emph
  {\bibinfo {title} {A Guide to First-Passage Processes}}}\ (\bibinfo
  {publisher} {Cambridge University Press},\ \bibinfo {year}
  {2001})\BibitemShut {NoStop}%
\bibitem [{\citenamefont {B{\'{e}}nichou}\ \emph {et~al.}(2015)\citenamefont
  {B{\'{e}}nichou}, \citenamefont {Gu{\'{e}}rin},\ and\ \citenamefont
  {Voituriez}}]{Bnichou2015}%
  \BibitemOpen
  \bibfield  {author} {\bibinfo {author} {\bibfnamefont {O.}~\bibnamefont
  {B{\'{e}}nichou}}, \bibinfo {author} {\bibfnamefont {T.}~\bibnamefont
  {Gu{\'{e}}rin}},\ and\ \bibinfo {author} {\bibfnamefont {R.}~\bibnamefont
  {Voituriez}},\ }\bibfield  {title} {\bibinfo {title} {Mean first-passage
  times in confined media: from markovian to non-markovian processes},\ }\href
  {https://doi.org/10.1088/1751-8113/48/16/163001} {\bibfield  {journal}
  {\bibinfo  {journal} {Journal of Physics A: Mathematical and Theoretical}\
  }\textbf {\bibinfo {volume} {48}},\ \bibinfo {pages} {163001} (\bibinfo
  {year} {2015})}\BibitemShut {NoStop}%
\bibitem [{\citenamefont {{Heller}}(1987)}]{1987PhRvA..35.1360H}%
  \BibitemOpen
  \bibfield  {author} {\bibinfo {author} {\bibfnamefont {E.~J.}\ \bibnamefont
  {{Heller}}},\ }\bibfield  {title} {\bibinfo {title} {{Quantum localization
  and the rate of exploration of phase space}},\ }\href
  {https://doi.org/10.1103/PhysRevA.35.1360} {\bibfield  {journal} {\bibinfo
  {journal} {\pra}\ }\textbf {\bibinfo {volume} {35}},\ \bibinfo {pages} {1360}
  (\bibinfo {year} {1987})}\BibitemShut {NoStop}%
\bibitem [{\citenamefont {Cohen}\ \emph {et~al.}(2016)\citenamefont {Cohen},
  \citenamefont {Yukalov},\ and\ \citenamefont {Ziegler}}]{cohen16}%
  \BibitemOpen
  \bibfield  {author} {\bibinfo {author} {\bibfnamefont {D.}~\bibnamefont
  {Cohen}}, \bibinfo {author} {\bibfnamefont {V.~I.}\ \bibnamefont {Yukalov}},\
  and\ \bibinfo {author} {\bibfnamefont {K.}~\bibnamefont {Ziegler}},\
  }\bibfield  {title} {\bibinfo {title} {Hilbert-space localization in closed
  quantum systems},\ }\href {https://doi.org/10.1103/PhysRevA.93.042101}
  {\bibfield  {journal} {\bibinfo  {journal} {Phys. Rev. A}\ }\textbf {\bibinfo
  {volume} {93}},\ \bibinfo {pages} {042101} (\bibinfo {year}
  {2016})}\BibitemShut {NoStop}%
\bibitem [{\citenamefont {{Berry}}(1984)}]{berry84}%
  \BibitemOpen
  \bibfield  {author} {\bibinfo {author} {\bibfnamefont {M.~V.}\ \bibnamefont
  {{Berry}}},\ }\bibfield  {title} {\bibinfo {title} {{Quantal Phase Factors
  Accompanying Adiabatic Changes}},\ }\href
  {https://doi.org/10.1098/rspa.1984.0023} {\bibfield  {journal} {\bibinfo
  {journal} {Proceedings of the Royal Society of London Series A}\ }\textbf
  {\bibinfo {volume} {392}},\ \bibinfo {pages} {45} (\bibinfo {year}
  {1984})}\BibitemShut {NoStop}%
\end{thebibliography}%

\end{document}